\documentclass[aps,pre,showpacs,twocolumn]{revtex4}
\usepackage{bm}
\usepackage{graphicx}
\bibstyle{approve.bib}
\usepackage{amssymb}
\usepackage{amsmath}
\usepackage{esint}
\usepackage{epstopdf}
\usepackage{color}
\usepackage{gensymb}
\usepackage{mathtools}
\usepackage{suffix}

\newcommand{\be}{\begin{equation}}
\newcommand{\ee}{\end{equation}}
\newcommand{\bea}{\begin{eqnarray}}
\newcommand{\eea}{\end{eqnarray}}

\newcommand{\br}{\mathbf{r}}

\newcommand{\e}{\varepsilon}

\newcommand{\tv}{\tilde{v}}
\newcommand{\tr}{\tilde{r}}
\newcommand{\td}{\tilde{d}}

\begin{document}

\title{Like-charge attraction and opposite-charge decomplexation between polymers and DNA molecules}

\author{Sahin Buyukdagli}
\address{Department of Physics, Bilkent University, Ankara 06800, Turkey}

\date{\small\it \today}

\begin{abstract}
We scrutinize the effect of polyvalent ions on polymer-DNA interactions. We extend a recently developed test charge theory [S. Buyukdagli \textit{et al.}, Phys. Rev. E   \textbf{94}, 042502 (2016)] to the case of a stiff polymer interacting with a DNA molecule in an electrolyte mixture. The theory accounts for one-loop level electrostatic correlation effects such as the ionic cloud deformation around the strongly charged DNA molecule as well as image-charge forces induced by the low DNA permittivity. Our model can reproduce and explain various characteristics of the experimental phase diagrams for polymer solutions. First, the addition of polyvalent cations to the electrolyte solution results in the attraction of the negatively charged polymer by the DNA molecule. The glue of the like-charge attraction is the enhanced shielding of the polymer charges by the dense counterion layer at the DNA surface. Secondly, through the shielding of the DNA-induced electrostatic potential, mono- and polyvalent cations of large concentration both suppress the like-charge attraction. Within the same formalism, we also predict a new opposite-charge repulsion effect between the DNA molecule and a positively charged polymer. In the presence of polyvalent anions such as sulfate or phosphate, their repulsion by the DNA charges leads to the charge screening deficiency of the region around the DNA molecule. This translates into a repulsive force that results in the decomplexation of the polymer from DNA. This opposite-charge repulsion phenomenon can be verified by current experiments and the underlying mechanism can be beneficial to gene therapeutic applications where the control over polymer-DNA interactions is the key factor.

\end{abstract}


\date{\today}
\maketitle

\section{Introduction}

DNA is the central building block of life. The biological functions of this anionic molecule are intimately related to its electrostatic coupling with the surrounding macromolecular charges. In this context, the \textit{like-charge attraction} phenomenon plays a capital role in DNA packing around histones~\cite{PodRev}, the condensation of polymer solutions~\cite{exp1,Levin,Netz2000,Cruz1,Cruz2,Lobaskin2003,Deserno,Naji1}, DNA adsorption on anionic substrates~\cite{Molina2014I,Molina2014II,PRE2016,Levin3}, and gene delivery techniques~\cite{Molina2014}.  The characterization of like-charge attraction is thus of paramount importance for understanding and controlling  in-vivo and in-vitro biological processes involving DNA molecules.

On the experimental side, it is well established that polyvalent counterions are the essential ingredient of like-charge polymer attraction~\cite{exp2,exp3,exp4,exp5,exp6,exp7,exp8,exp9,Cruz}. More precisely, these experiments showed that the occurrence of like-charged polymer condensation at physiological salt densities necessitates the presence of trivalent counterions in the electrolyte solution.  It should be noted that the polyvalent cation-induced like-charge attraction cannot be explained from the perspective of mean-field (MF) electrostatics; regardless of the electrolyte composition and the ionic valency, the MF theory always predicts repulsion between similar charges.  In order to interpret this seemingly counterintuitive phenomenon, a beyond-MF formulation of the problem including charge correlations is needed. In this direction, previous theories have considered various mechanisms such as counterion binding~\cite{Cruz1,Cruz2,Cruz,Ha97,Ha99,goles,Levin2,Muthu}, attractive van der Waals interactions~\cite{Podg98}, and Wigner crystal formation occurring in the electrostatic strong-coupling regime~\cite{Shkl99,jensen97}. \textcolor{black}{More precisely, in Refs.~\cite{Cruz1,Cruz2} and~\cite{Cruz}, the authors investigated like-charged polymer attraction within a Landau-Ginzburg-like approach including the microscopic details of the Flory-Huggins free energy. In order to account for electrostatic correlation effects explicitly, approximative electrostatic theories were used in Refs.~\cite{ Ha97,Ha99,goles,Levin2,Muthu,Shkl99,jensen97}. Namely Refs.~\cite{Ha97,Ha99,goles} opted for gaussian path-integral techniques to consider electrostatic many-body effects. Then, a free energetic approach that includes charge correlations and polymer fluctuations was used by Muthukumar in Ref.~\cite{Muthu}. In Ref.~\cite{Levin2}, Arenzon et al. formulated the polymer attraction problem within a discrete counterion binding model. Shklovskii considered the effect of polyvalent cations on DNA condensation from the perspective of the strong coupling theory~\cite{Shkl99}. Finally, in Ref.~\cite{jensen97}, a rigid rod model in contact with multivalent counterions was simulated and also theoretically investigated within the one-component plasma approximation.}

\textcolor{black}{The large diversity of the above-mentioned approaches to like-charge polymer condensation clearly indicates the need to build-up a systematic theoretical framework for the investigation of this problem. In this article, we scrutinize the mechanism driving like-charge attraction from a bottom-up approach. Within the field-theoretic one-loop formulation of charged liquids, we calculate the electrostatic free energy of the polymer-DNA complex by including consistently the microscopic details of the surrounding electrolyte mixture.} We find that in the presence of polyvalent counterions, like-charge attraction emerges naturally as a consequence of the non-uniform shielding of the polymer charges in the inhomogeneous screening background set-up by ion-DNA interactions. Within the same formalism, we also predict a new opposite charge repulsion mechanism. Namely, we consider the opposite case of the DNA interacting with a positively charged polymer. Upon the addition of polyvalent anions to the solution, the anion depletion from the DNA surface translates into a repulsive force that results in the decomplexation of the oppositely charged polymer from the DNA molecule.

This article introduces the first formulation of polymer-DNA interactions within the one-loop (1l) theory of charge correlations.  The 1l formalism has the advantage of being a controlled approximation that also covers the electrostatic MF regime of monovalent electrolytes~\cite{Netz2000}. The latter point is crucial for the present problem where we intend to scrutinize the deviation from the MF behaviour upon the addition of polyvalent ions to a monovalent electrolyte solution. In the inclusion of 1l-level correlations to polymer-DNA interactions, a first technical complication arises from the breaking of the cylindrical symmetry by the linear polymer. At this point, we make use of the \textit{test charge approach} that we have recently developed in Ref.~\cite{PRE2016}. Therein, the approach was applied to the case of a polymer interacting with a like-charged plane wall in order to explain the adsorption of DNA molecules onto negatively charged lipid membranes~\cite{Molina2014I}. In the present problem, the test charge approach allows us to recover the cylindrical symmetry dictated by the geometry of the DNA molecule. The additional difficulty associated with the 1l theory stems from the electrolyte composition and the cylindrical geometry of the problem. Indeed, in the presence of an electrolyte mixture including different species of mono- and polyvalent ions, the radial electrostatic Green's equation cannot be solved analytically. We overcome this difficulty by adapting a numerical solution scheme developed in Refs.~\cite{jcp2,Buyuk2014II} to the cylindrical geometry of the DNA molecule. In the articles mentioned above, the quantitative reliability of the underlying 1l-theory and its validity domain were determined by comparison with Monte-Carlo simulations for polyvalent ion partition in nanopores and also against experimental ionic conductivity data in subnanometer pores. \textcolor{black}{At this point, we also emphasize that in contrast with the approximations of the previous theoretical approaches in Refs.~\cite{ Ha97,Ha99,goles,Levin2,Muthu}, our 1l test charge approximation can be systematically improved in terms of the consideration of charge correlations and the polymer charge density. We note that this point was shown and the correction terms were formally derived in our recent article Ref.~\cite{PRE2016}.}

Our article is organised as follows. In Sec.~\ref{subsec1},  we extend the test charge theory of Ref.~\cite{PRE2016} to mixed electrolytes. In Sec.~\ref{subsec2}, using the test charge approach, we calculate the interaction potential of the polymer-DNA complex in an electrolyte mixture. Sec.~\ref{at} is devoted to the characterization of the mechanism driving the like-charge polymer-DNA attraction. In qualitative agreement with experiments on polymer solutions~\cite{exp2,exp3,exp4,exp5,exp6,exp7,exp8,exp9,Cruz}, we find that like-charge attraction emerges upon the addition of multivalent cations but the effect is suppressed by monovalent cations or polyvalent cations of large concentration. Finally, in Sec.~\ref{op1}, we consider the interaction of a positively-charged polymer with the DNA molecule. Therein, the addition of polyvalent anions to the electrolyte solution is shown to induce an opposite charge repulsion effect resulting in the decomplexation of the polymer from the DNA surface. We summarize our main results and discuss the limitations and possible extensions of our formalism in the Conclusion part.

\section{Theory}
\label{th}

In this part, we introduce the theoretical framework for the characterization of polymer-DNA interactions in electrolyte mixtures. In Sec.~\ref{subsec1}, we develop the \textit{test charge approach} that allows to calculate the grand potential of a weakly charged molecule interacting with a macromolecule of arbitrary charge strength. This derivation is a generalisation of the approach introduced in Ref.~\cite{PRE2016} for symmetric electrolytes to the case of electrolyte mixtures composed of an arbitrary number of ionic species. The test charge approach is applied in Sec.~\ref{subsec2} to the specific geometry of a linear polymer interacting with a cylindrical DNA molecule. The derivation of the polymer grand potential requires the solution of a non-uniformly screened Debye-H\"{u}ckel (DH) equation. In the same section, we explain in detail an inversion scheme that allows to solve this equation by iteration.

\subsection{Grand potential of a test charge interacting with a macromolecule}
\label{subsec1}

We derive here the grand potential of a test charge interacting with a charged macromolecule. Both molecules are immersed in an electrolyte mixture. Our starting point is the MF grand potential of an electrolyte mixture including as well charged bodies treated as fixed charges of arbitrary geometry,
\bea\label{1}
\Omega&=&-\frac{k_BT}{2e^2}\int\mathrm{d}\br\e(\br)\left[\nabla\phi(\br)\right]^2+\int\mathrm{d}\br\sigma(\br)\phi(\br)\nonumber\\
&&-\sum_{i=1}^p\rho_{bi}\int\mathrm{d}\br e^{-V_i(\br)-q_i\phi(\br)}.
\eea
The first term on the r.h.s. of Eq.~(\ref{1}) is the electrostatic free energy of the solvent. Therein, $k_BT$ stands for the thermal energy with $T=300$ K the ambient temperature, $e$ is the unit charge,  and $\e(\br)$ the dielectric permittivity profile. Furthermore, $\phi(\br)$ stands for the electrostatic potential induced by the fixed charges of density $\sigma(\br)$ taken into account by the second integral term. Finally, the third term accounts for the presence of  $p$ mobile ion species. The ions of the species $i$ with the bulk density $\rho_{bi}$ and valency $q_i$ experience the steric potential $V_i(\br)$ accounting for their depletion from the location of the macromolecular charges. We finally note that in the bulk, i.e. infinitely far from the macromolecular charges, the electroneutrality condition reads
\be\label{1II}
\sum_{i=1}^p\rho_{bi}q_i=0.
\ee

The MF-level electrostatic potential is obtained by optimising Eq.~(\ref{1}) with respect to the potential $\phi(\br)$, i.e. $\delta\Omega/\delta\phi(\br)=0$. This yields the PB equation
\be
\label{2}
\nabla\e(\br)\nabla\phi(\br)+\frac{e^2}{k_BT}\sum_{i=1}^p\rho_{bi}q_ie^{-V_i(\br)-q_i\phi(\br)}=-\frac{e^2}{k_BT}\sigma(\br).
\ee
From now on, we will consider the specific case of a \textit{weakly charged body} of density $\sigma_p(\br)$ that will be treated as a \textit{test charge}, interacting with a macromolecule with charge density $\sigma_d(\br)$ of arbitrary strength. Eqs.~(\ref{1}) and~(\ref{2}) will be expanded in terms of the test charge density. To this aim, we express the total fixed charge density and electrostatic potential as 
\bea
\label{3}
\sigma(\br)&=&\sigma_d(\br)+\lambda\sigma_p(\br)\\
\label{4}
\phi(\br)&=&\phi_d(\br)+\lambda\phi_p(\br),
\eea
where $\phi_d(\br)$ and $\phi_p(\br)$ are the electrostatic potential components induced by the charged macromolecule and the test charge, respectively. Moreover, $\lambda$ is a perturbative coefficient that will be set to unity at the end of the expansion.  Next, we inject Eqs.~(\ref{3}) and~(\ref{4}) into the PB Eq.~(\ref{2}) and expand the result at the linear order in the parameter $\lambda$. This yields the equations of state associated with the charged macromolecule and the test charge,
\bea\label{5}
&&\nabla\e(\br)\nabla\phi_d(\br)+\frac{e^2}{k_BT}\sum_{i=1}^pq_in_i(\br)=-\frac{e^2}{k_BT}\sigma_d(\br),\\
\label{6}
&&\left\{\nabla\e(\br)\nabla-\frac{e^2}{k_BT}\sum_{i=1}^pq_i^2n_i(\br)\right\}\phi_p(\br)=-\frac{e^2}{k_BT}\sigma_p(\br),\nonumber\\
\eea
where we introduced the MF-level ionic number density 
\be\label{7}
n_i(\br)=\rho_{bi}e^{-V_i(\br)-q_i\phi_d(\br)}.
\ee

First, we need to solve Eq.~(\ref{6}). To this aim, we introduce the kernel operator defined as the inverse of the Green's function $v(\br,\br')$,
\be\label{8}
v^{-1}(\br,\br')=\left\{-\frac{k_BT}{e^2}\nabla\e(\br)\nabla+\sum_{i=1}^pq_i^2n_i(\br)\right\}\delta(\br-\br').
\ee
This allows to express Eq.~(\ref{6}) in the form
\be
\label{9}
\int\mathrm{d}\br'v^{-1}(\br,\br')\phi_p(\br')=\sigma_p(\br).
\ee
Multiplying now Eq.~(\ref{9}) by the Green's function $v(\br'',\br)$, integrating once, and using the definition of the Green's function,
\be\label{10}
\int\mathrm{d}\br''\;v^{-1}(\br,\br'')v(\br'',\br')=\delta(\br-\br'),
\ee
the electrostatic potential associated with the test charge finally takes the form
\be\label{11}
\phi_p(\br)=\int\mathrm{d}\br'v(\br,\br')\sigma_p(\br').
\ee
Now, we inject the relations~(\ref{3}) and~(\ref{4}) into the grand potential~(\ref{1}) and expand the result at the quadratic order in $\lambda$. Using Eqs.~(\ref{6}) and~(\ref{8}) to simplify the result, after some easy algebra, the total grand potential splits into the components associated with the macromolecule and the test charge, $\Omega=\Omega_d+\Omega_p$. In this equation, the macromolecular grand potential $\Omega_d$ corresponds to Eq.~(\ref{1}) with the total potential $\phi(\br)$ and fixed charge density $\sigma(\br)$ replaced by the macromolecular potential $\phi_d(\br)$ and charge $\sigma_d(\br)$, respectively. The grand potential of the test charge in turn reads 
\bea
\label{12}
\Omega_p&=&-\frac{1}{2}\int\mathrm{d}\br\mathrm{d}\br'\phi_p(\br)v^{-1}(\br,\br')\phi_p(\br')\\
&&+\int\mathrm{d}\br\;\sigma_p(\br)\left[\phi_d(\br)+\phi_p(\br)\right].\nonumber
\eea
Inserting the integral expression~(\ref{11}) into Eq.~(\ref{12}), the grand potential of the test charge finally takes the form
\be
\label{13}
\Omega_p=\frac{1}{2}\int\mathrm{d}\br\mathrm{d}\br'\sigma_p(\br)v(\br,\br')\sigma_p(\br')+\int\mathrm{d}\br\sigma_p(\br)\phi_d(\br).
\ee
At this point, we note that the physically relevant quantity that determines the nature of the interaction between the test charge and the macromolecule is the difference between the total grand potential~(\ref{13}) and its bulk limit. Since the macromolecular potential $\phi_d(\br)$ vanishes in the bulk, the grand potential difference is given by
\bea
\label{15}
\Delta\Omega_p&=&\frac{1}{2}\int\mathrm{d}\br\mathrm{d}\br'\sigma_p(\br)\left[v(\br,\br')-v_b(\br-\br')\right]\sigma_p(\br')\nonumber\\
&&+\int\mathrm{d}\br\sigma_p(\br)\phi_d(\br).
\eea
In Eq.~(\ref{15}), we introduced the spherically symmetric bulk Green's function $v_b(\br-\br')=\ell_Be^{-\kappa_b|\br-\br'|}/|\br-\br'|$, with the DH screening parameter $\kappa_b^2=4\pi\ell_B\sum_{i=1}^p\rho_{bi}q_i^2$ and the Bjerrum length $\ell_B\approx 7$ {\AA}. 

Finally, one notes that the evaluation of Eq.~(\ref{15}) requires the electrostatic potential $\phi_d(\br)$ and the Green's function $v(\br,\br')$. The former is obtained from the solution of Eq.~(\ref{5}) by a standard numerical discretization procedure. In order to obtain the Green's function, we have to derive first the corresponding kernel equation. To this aim, we multiply both sides of Eq.~(\ref{8}) by the Green's function and integrate once. Using the relation~(\ref{10}), one obtains 
\be
\label{14}
\left\{\nabla\e(\br)\cdot\nabla-\frac{e^2}{k_BT}\sum_{i=1}^pq_i^2n_i(\br)\right\}v(\br,\br')=-\frac{e^2}{k_BT}\delta(\br-\br').
\ee
We note that Eq.~(\ref{14}) together with Eq.~(\ref{7}) corresponds to the 1l level kernel equation for the electrostatic propagator~\cite{Netz2000,jcp2,Buyuk2014II}. The numerical scheme for the solution of Eq.~(\ref{14}) will be explained in Sec.~\ref{subsec2}. 

\subsection{Polymer interacting with a ds-DNA molecule}
\label{subsec2}

\begin{figure}
\includegraphics[width=1.\linewidth]{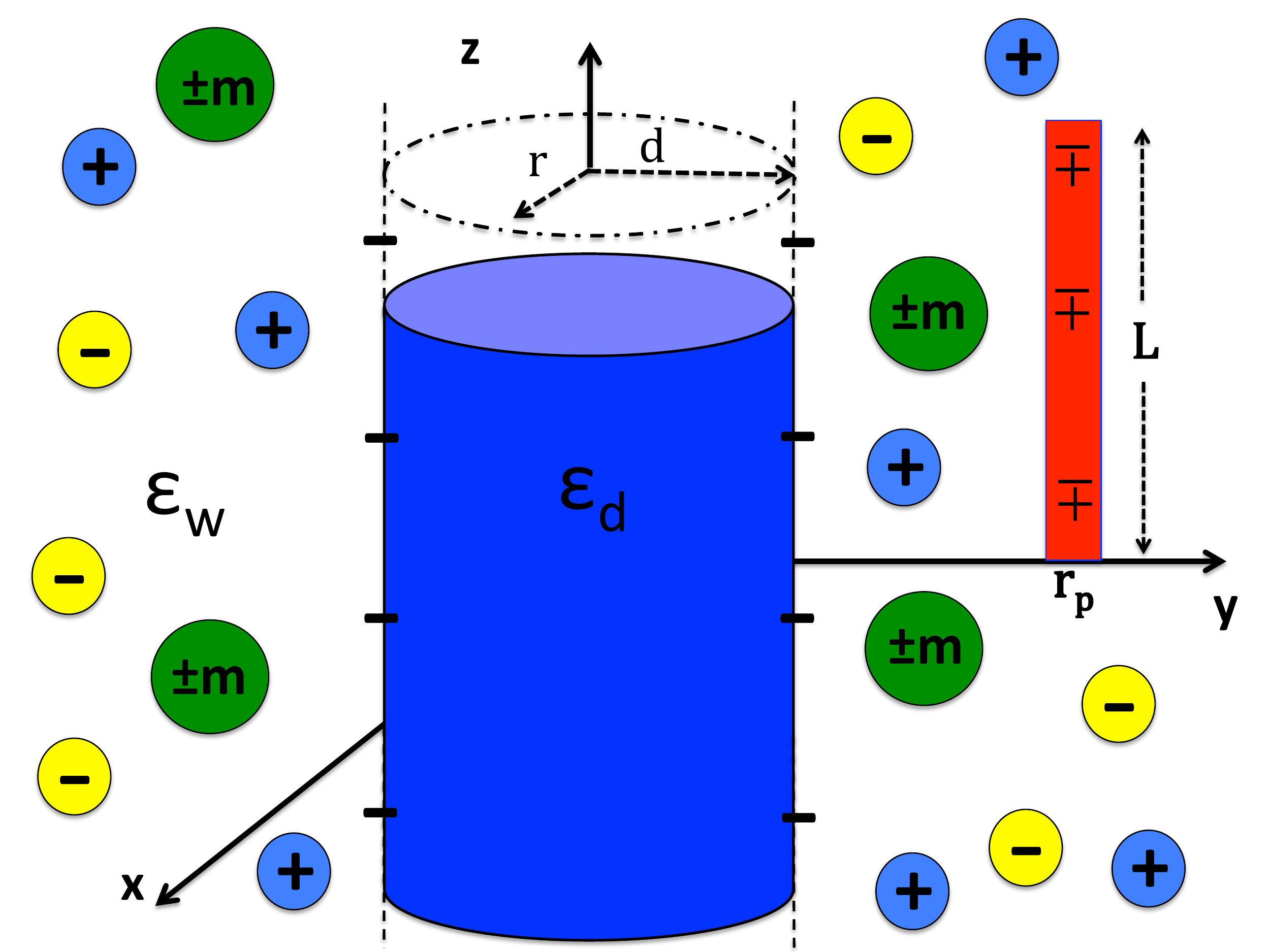}
\caption{(Color online) Schematic depiction of the electrolyte mixture including a DNA molecule (blue) and a polymer (red). The negative DNA charge density has magnitude $\sigma_d=0.4$ $e/\mathrm{nm}^2$~\cite{dnaden}, with the DNA radius $d=1$ nm and dielectric permittivity $\e_d=2$. The polyelectrolyte has length $L$ and charge density $\tau$ of positive or negative sign. The electrolyte mixture contains the monovalent ions $\mbox{Na}^+$ and $\mbox{Cl}^-$, and polyvalent ions with valency $m$ of positive ($+m$) or negative ($-m$) net charge.}
\label{fig1}
\end{figure}

\subsubsection{Derivation of the polymer grand potential}

We consider now the specific case of a polymer interacting with a double-stranded DNA (ds-DNA) molecule in an electrolyte mixture. The charge configuration of the system is depicted in Fig.~\ref{fig1}. The DNA molecule (blue) and the charged polymer (red) are immersed in an electrolyte mixture including monovalent ions $\mbox{Na}^+$ and $\mbox{Cl}^-$, and polyvalent ions of negative or positive net charge and valency $m$.  Our choice of the polyelectrolyte model is motivated by previous experiments where macromolecular conformations were shown to play no qualitative role in like charge aggregation~\cite{exp4}. More precisely, single stranded DNA (ss-DNA) and ds-DNA molecules, nucleosome core particles, and chromatin fragments characterized by radically different charge configurations were shown to exhibit qualitatively the same precipitation characteristics.  Based on this observation, we neglect polymer fluctuations and model the interacting polyelectrolytes as stiff macromolecules. Namely, the negatively charged DNA molecule corresponds to an infinitely long cylinder of radius $d$, whose axis of symmetry coincides with the $z-$axis. The polymer is modelled as a stiff rod of length $L$ and charge density $\tau$. It is oriented parallel with the DNA molecule and located at the radial distance $r_p$ from the $z-$axis. We will consider both the case of negatively charged ($\tau<0$) and positively charged ($\tau>0$) polymers.

In the charge configuration of Fig.~\ref{fig1}, the polymer and DNA charge densities are respectively given by
\bea
\label{16}
\sigma_p(\br)&=&\frac{\tau}{r}\delta(r-r_p)\delta(\varphi-\varphi_p)\theta(z)\theta(L-z)\\
\label{17}
\sigma_d(\br)&=&-\sigma_d\delta(r-d).
\eea
Due to the cylindrical symmetry, the potential induced by the DNA molecule depends solely on the radial distance from the DNA axis, i.e. $\phi_d(\br)=\phi_d(r)$. Moreover, within the same symmetry, one can Fourier-expand the Green's function as
\be\label{18}
v(\br,\br')=\sum_{m=-\infty}^{+\infty} e^{im(\varphi-\varphi')}\int_{-\infty}^{+\infty}\frac{\mathrm{d}k}{4\pi^2}e^{ik(z-z')}\tv_m(r,r';k).
\ee
Injecting Eqs.~(\ref{16})-(\ref{18}) into Eq.~(\ref{15}), the polymer grand potential takes the form
\be\label{19}
\Delta\Omega_p(r_p)=\Omega_{mf}(r_p)+\Delta\Omega_s(r_p),
\ee
with the MF-level interaction term
\be
\label{20}
\Omega_{mf}(r_p)=L\tau\phi_d(r_p),
\ee
and the polymer self-energy
\bea\label{21}
\Delta\Omega_s(r_p)&=&\frac{L\tau^2}{4\pi}\sum_{m=-\infty}^{+\infty}\int_{-\infty}^{+\infty}\mathrm{d}k\frac{2\sin^2(kL/2)}{\pi k^2L}\\
&&\hspace{1cm}\times\left[\tv_m(r_p,r_p;k)-\tv_{b,m}(r_p,r_p;k)\right].\nonumber
\eea
In this work, we will consider the biologically relevant regime of long polymers and take the limit $L\to\infty$. In this limit, the sinusoidal function in the integral tends to a Dirac delta distribution and the polymer self-energy reduces to
\be\label{21II}
\Delta\Omega_s(r_p)=\frac{L\tau^2}{4\pi}\sum_{m=-\infty}^{+\infty}\left[\tv_m(r_p,r_p;0)-\tv_{b,m}(r_p,r_p;0)\right].
\ee

\subsubsection{Solving the electrostatic kernel equation~(\ref{14})}

The evaluation of the potential components~(\ref{20}) and~(\ref{21II}) requires the knowledge of the potentials $\phi_d(r)$ and $\tv_m(r,r';k)$. In the cylindrically symmetric charge distribution~(\ref{17}), the differential equations~(\ref{5}) and~(\ref{14}) solved by these potentials read
\bea\label{22}
&&\frac{k_BT}{e^2}\frac{1}{r}\partial_r\left[r\e(r)\partial_r\phi_d(r)\right]+\sum_{i=1}^pq_in_i(r)=\sigma_d\delta(r-d),\nonumber\\
&&\\
\label{24}
&&\left\{\frac{1}{r}\partial_rr\e(r)\partial_r-\e(r)\left[\frac{m^2}{r^2}+k^2+\kappa^2(r)\right]\right\}\tv_m(r,r';k)\nonumber\\
&&=-\frac{e^2}{k_BT}\frac{1}{r}\delta(r-r'),
\eea
with the ionic number density  
\be\label{23}
n_i(r)=\rho_{bi}e^{-q_i\phi_d(r)}\theta(r-d),
\ee
the dielectric permittivity profile
\be\label{27II}
\e(r)=\e_w\theta(r-d)+\e_d\theta(d-r),
\ee
and the local screening function
\be
\label{25}
\kappa^2(r)=4\pi\ell_B\sum_{i=1}^pq^2_in_i(r).
\ee
In Eq.~(\ref{27II}), the dielectric permittivities of the electrolyte and the DNA molecule are $\e_w=80$ and $\e_d=2$, respectively. Using Gauss' law $\phi'(d)=4\pi\ell_B\sigma_d$, Eq.~(\ref{22}) can be numerically solved by discretisation. However, the solution of Eq.~(\ref{24}) is non-trivial. Indeed, this kernel equation is a non-uniformly screened DH equation that presents no analytical solution. Hence, we will first transform this differential equation into a numerically tractable integral relation. We present next this solution scheme.

Our strategy consists in solving Eq.~(\ref{24}) by iteration around the weak-coupling (WC) Debye-H\"{u}ckel (DH) Green's function $\tv_{0,m}(r,r';k)$. To this aim, we introduce the corresponding kernel equation
\bea
\label{26}
&&\left\{\frac{1}{r}\partial_rr\e(r)\partial_r-\e(r)\left[\frac{m^2}{r^2}+k^2+\kappa_b^2(r)\right]\right\}\tv_{0,m}(r,r';k)\nonumber\\
&&=-\frac{e^2}{k_BT}\frac{1}{r}\delta(r-r'),
\eea
with the piecewise screening function
\be
\label{27}
\kappa_b(r)=\kappa_b\theta(r-d).
\ee
Using the Fourier transform of Eq.~(\ref{10}),
\be\label{28}
\int_0^{\infty}\mathrm{d}r''r''\tv_{0,m}^{-1}(r,r'';k)\tv_{0,m}(r'',r';k)=\frac{1}{r}\delta(r-r'),
\ee
one can show that the DH kernel associated with Eq.~(\ref{26}) is
\bea\label{29}
\tv^{-1}_{0,m}(r,r';k)&=&-\frac{k_BT}{e^2}\left\{\frac{1}{r}\partial_rr\e(r)\partial_r\right.\\
&&\hspace{8mm}\left.-\e(r)\left[\frac{m^2}{r^2}+k^2+\kappa_b^2(r)\right]\right\}\frac{\delta(r-r')}{r}.\nonumber
\eea
Next we express Eq.~(\ref{24}) in a form similar to the DH Eq.~(\ref{26}),
\bea
\label{30}
&&\left\{\frac{1}{r}\partial_rr\e(r)\partial_r-\e(r)\left[\frac{m^2}{r^2}+k^2+\kappa_b^2(r)\right]\right\}\tv_m(r,r';k)\nonumber\\
&&=-\frac{e^2}{k_BT}\frac{1}{r}\delta(r-r')+\e(r)\left[\kappa^2(r)-\kappa_b^2(r)\right]\tv_m(r,r';k).\nonumber\\
\eea
In terms of the DH kernel operator~(\ref{29}), Eq.~(\ref{30}) can be recast as
\bea
\label{31}
&&-\frac{e^2}{k_BT}\int_0^{\infty}\mathrm{d}r_2r_2\tv_{0,m}^{-1}(r_1,r_2;k)\tv_m(r_2,r';k)\\
&&=-\frac{e^2}{k_BT}\frac{1}{r_1}\delta(r_1-r')\nonumber\\
&&\hspace{5mm}+\e(r_1)\left[\kappa^2(r_1)-\kappa_b^2(r_1)\right]\tv_m(r_1,r';k),\nonumber
\eea
where we made the substitution $r\to r_1$. Multiplying both sides of Eq.~(\ref{31}) by  $r_1\tv_{0,m}(r,r_1;k)$, integrating once, and using Eq.(\ref{28}), the kernel equation~(\ref{24}) finally takes the integral form
\bea
\label{32}
\tv_m(r,r';k)&=&\tv_{0,m}(r,r';k)\\
&&+\int_d^\infty\mathrm{d}r''r''\tv_{0,m}(r,r'';k)\delta n(r'')\tv_m(r'',r';k),\nonumber
\eea
with the density excess function 
\be\label{33}
\delta n(r)=\sum_{i=1}^p\rho_{bi}q_i^2\left[1-e^{-q_i\phi_d(r)}\right]\theta(r-d).
\ee

The solution of the integral Eq.~(\ref{32}) requires the knowledge of the DH Green's function solving the DH Eq.~(\ref{26}). In the present problem where the charge sources are located outside the cylinder, i.e. $r'>d$, the homogeneous solution to Eq.~(\ref{26}) is given by the linear combination of the modified Bessel functions~\cite{math},
\bea\label{34}
\tv_{0,m}(r,r';k)&=&C_1I_m(kr)\theta(d-r)\\
&&+\left[C_2I_m(pr)+C_3K_m(pr)\right]\theta(r-d)\theta(r'-r)\nonumber\\
&&+C_4K_m(pr)\theta(r-d)\theta(r-r'),\nonumber
\eea
where we introduced the parameter $p=\sqrt{k^2+\kappa_b^2}$. The integration constants $\left\{C_i\right\}_{1\leq i\leq4}$ are to be determined from the boundary conditions
\bea\label{35}
&&\lim_{r\to d^+}\tv_{0,m}(r,r';k)=\lim_{r\to d^-}\tv_{0,m}(r,r';k)\\
\label{36}
&&\lim_{r\to r'^+}\tv_{0,m}(r,r';k)=\lim_{r\to r'^-}\tv_{0,m}(r,r';k)\\
\label{37}
&&\lim_{r\to d^+}\e(r)\partial_r\tv_{0,m}(r,r';k)=\lim_{r\to d^-}\e(r)\partial_r\tv_{0,m}(r,r';k)\\
\label{38}
&&\lim_{r\to r'^+}\partial_r\tv_{0,m}(r,r';k)-\lim_{r\to r'^-}\partial_r\tv_{0,m}(r,r';k)=-\frac{4\pi\ell_B}{r'}.\nonumber\\
\eea
Imposing the conditions~(\ref{35})-(\ref{38}) to Eq.~(\ref{34}), the Green's function becomes for $r>d$
\bea\label{39}
\tv_{0,m}(r,r';k)&=&4\pi\ell_B\left[I_m(pr_<)K_m(pr_>)\right.\\
&&\hspace{9mm}\left.+F_m(k)K_m(pr_<)K_m(pr_>)\right],\nonumber
\eea
where we introduced the variables 
\bea\label{39II}
r_<&=&\mathrm{min}(r,r') \\
r_>&=&\mathrm{max}(r,r'), 
\eea
and the auxiliary function associated with the dielectric cylinder
\bea\label{40}
F_m(k)=\frac{\e_w\rho I_m(kd)I'_m(\rho d)-\e_dk I_m(\rho d)I'_m(k d)}{\e_dk K_m(\rho d)I'_m(k d)-\e_w\rho I_m(k d)K'_m(\rho d)}.
\eea
From Eq.~(\ref{39}), the bulk part of the Fourier-transformed Green's function used in Eqs.~(\ref{21}) and~(\ref{21II}) follows as
\be
\label{40II}
\tv_{b,m}(r,r';k)=4\pi\ell_BI_m(pr_<)K_m(pr_>).
\ee
With the knowledge of the average potential $\phi_d(r)$ present in Eq.~(\ref{33}) and the DH Green's function~(\ref{39}), the integral equation~(\ref{32}) can be solved by iteration. At the first iterative step, we evaluate the integral in Eq.~(\ref{32}) by replacing the Green's function $\tv_m(r,r';k)$ with its WC limit $\tv_{0,m}(r,r';k)$. The output function $\tv_m(r,r';k)$ is inserted at the next iterative step into the integral and this cycle is continued until numerical convergence is achieved.

\section{Anionic polymer-DNA attraction}
\label{at}
\begin{figure}
\includegraphics[width=1.1\linewidth]{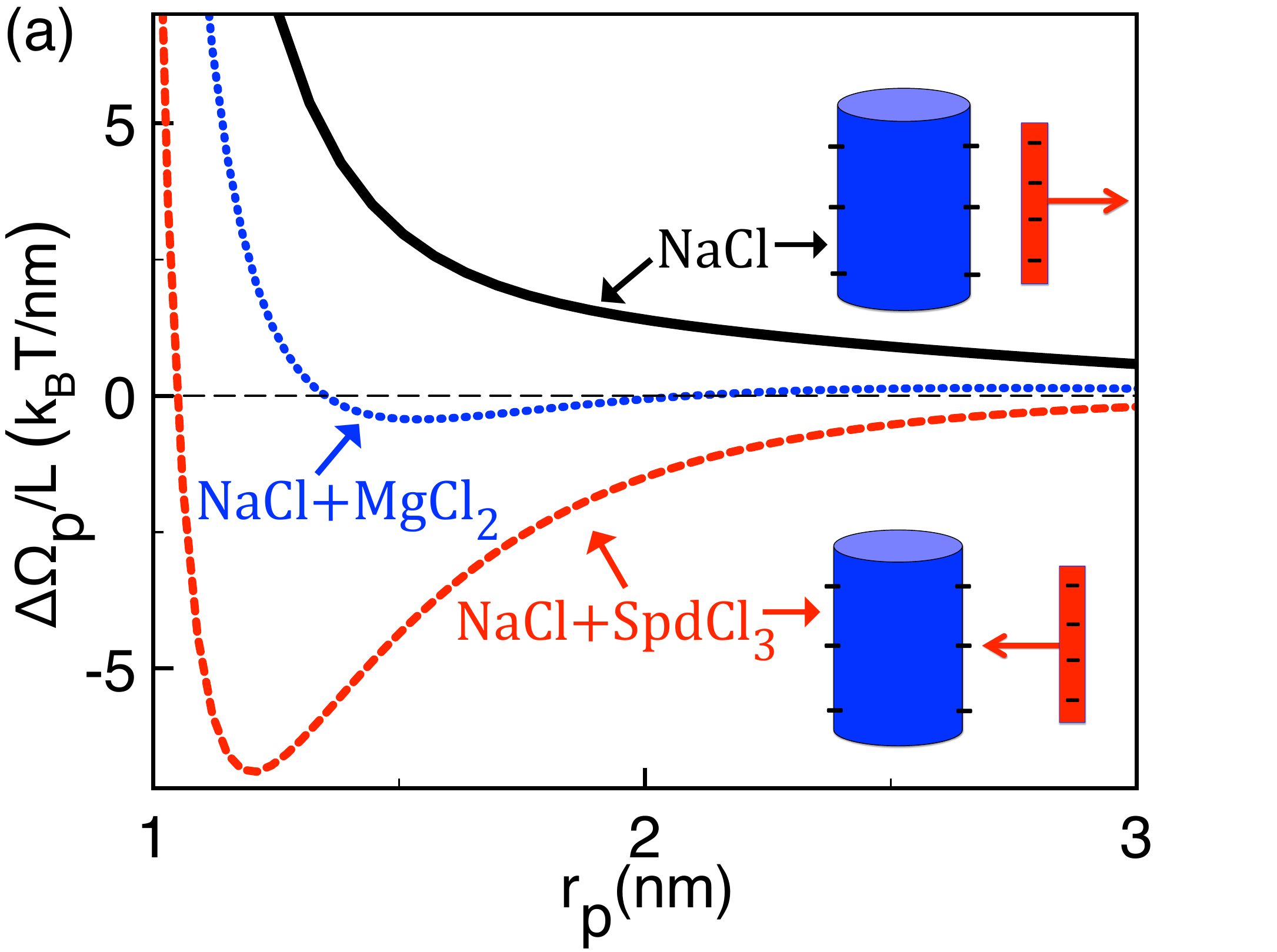}
\includegraphics[width=1.1\linewidth]{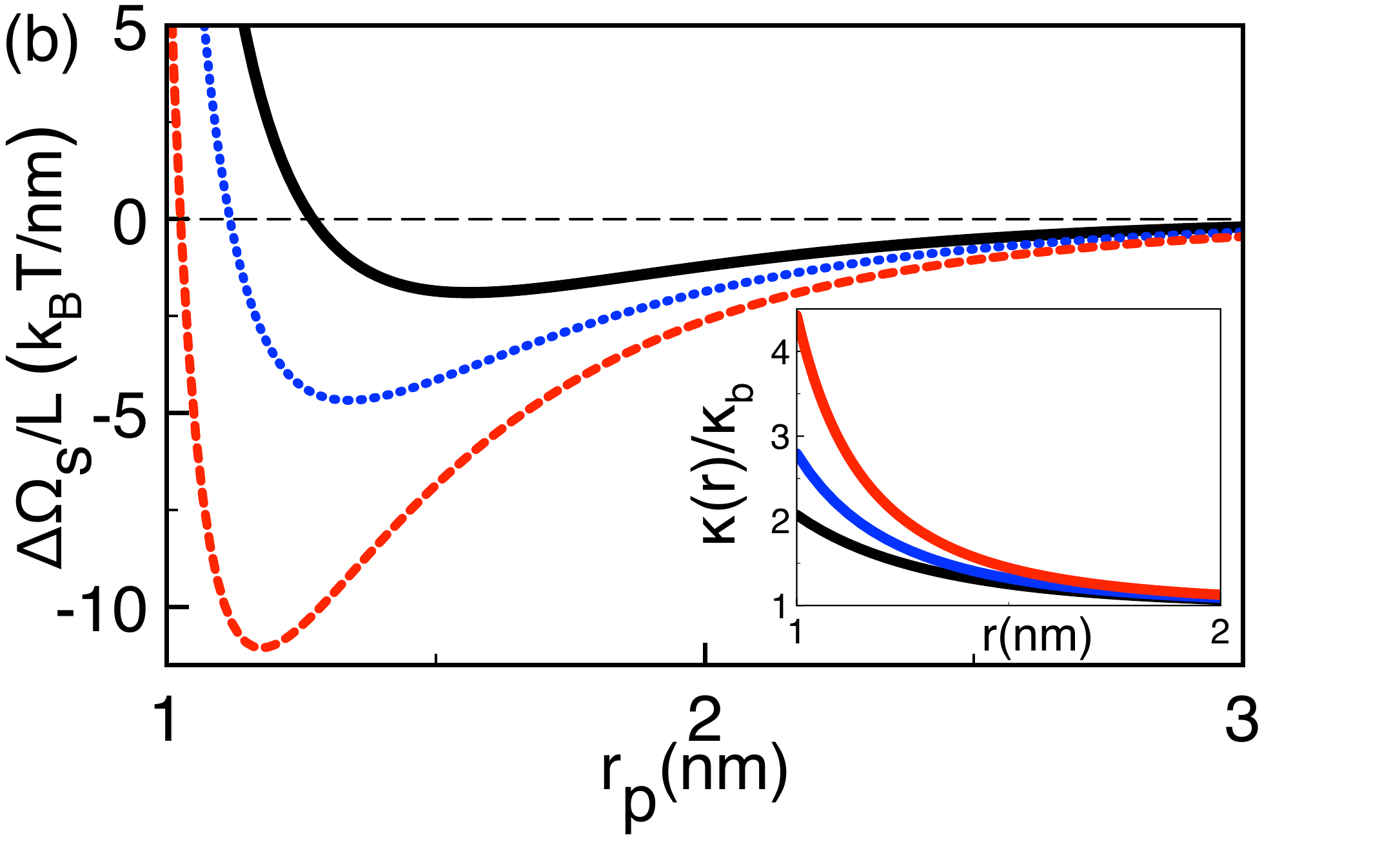}
\caption{(Color online) (a) Polymer grand potential $\Delta\Omega_p(r_p)=\Delta\Omega_s(r_p)+\Omega_{mf}(r_p)$, (b) self energy $\Delta\Omega_s(r_p)$ of Eq.~(\ref{21II}) (main plot), and rescaled local screening function from Eq.~(\ref{25})  (inset). The negatively charged polymer has charge density $\tau=-0.5$ $e/${\AA}.  Electrolyte composition: $\mbox{NaCl}$ (black curves), $\mbox{NaCl}+\mbox{MgCl}_2$ (blue curves) with $\mbox{Mg}^{2+}$ density $\rho_{b2+}=0.01$ M, and $\mbox{NaCl}+\mbox{SpdCl}_3$ (red curves) with $\mbox{Spd}^{3+}$ density $\rho_{b3+}=0.01$ M. $\mbox{Na}^+$ density is $\rho_{b+}=0.1$ M in all curves. The remaining parameters are given in the caption of Fig.~\ref{fig1}.}
\label{fig2}
\end{figure}

In this section, we scrutinize the interaction between the DNA molecule and a polyelectrolyte with negative charge density $\tau<0$. In addition to the monovalent sodium ($\mbox{Na}^+$) and chloride ($\mbox{Cl}^-$) ions, the electrolyte mixture $\mbox{NaCl}+\mbox{XCl}_m$ contains multivalent cations $\mbox{X}^{m+}$ that will be taken as divalent magnesium $\mbox{Mg}^{2+}$ or trivalent spermidine $\mbox{Spd}^{3+}$. In the following, the bulk mono-and polyvalent cation densities of the $\mbox{NaCl}+\mbox{XCl}_m$ mixture will be varied while the \textcolor{black}{bulk} anion density will be calculated according to the electroneutrality condition~(\ref{1II}),  i.e. $\rho_{b-}=\rho_{b+}+m\rho_{bm+}$. In Sec.~\ref{sec1}, the presence of polyvalent cations are shown to result in the attraction of the like-charge polymer by the DNA molecule. Therein, the electrostatic mechanism driving the like-charge polymer attraction is scrutinized in detail. In Sec.~\ref{sec1II}, we present an analytical WC expression for the polymer grand potential. This closed form expression provides analytical insight into the like-charge polymer attraction. The effect of monovalent cations and polymer charge strength on the like-charge attraction is considered in Secs.~\ref{sec2} and~\ref{sec3}, respectively. In these sections, we also compare the emerging interaction picture with the experimental phase diagrams of like-charged polymer solutions in electrolyte mixtures.

\subsection{Like-charge polymer-DNA attraction mechanism}
\label{sec1}

We consider here the like-charge polymer-DNA attraction effect. The grand potential profile of the negatively charged polymer is displayed in Fig.~\ref{fig2}(a). The polymer charge density is set to the value $\tau=-0.5$ $e/${\AA} located in the density range of single-stranded DNA molecules. In the monovalent $\mbox{NaCl}$ solution (black curve), the polymer-DNA interaction is purely repulsive. This is due to the combination of the MF-level like-charge repulsion brought by Eq.~(\ref{20}) and the polymer-image interactions included in the self energy~(\ref{21II}). The image-charge effect is induced by the low DNA permittivity. Adding divalent $\mbox{Mg}^{2+}$ cations to the solution (blue curve), an attractive potential minimum emerges. However, the corresponding polymer attraction is weak since the depth of the potential well is a fraction of the thermal energy,  $\Delta\Omega_p/L\simeq-0.4\;k_BT/\mathrm{nm}$. Replacing $\mbox{Mg}^{2+}$ ions by trivalent $\mbox{Spd}^{3+}$ molecules (red curve), the interaction becomes strongly attractive as the potential minimum drops to the significant value $\Delta\Omega_p/L\simeq-7\;k_BT/\mathrm{nm}$.

The mechanism driving the like-charge polyelectrolyte attraction can be explained by the behaviour of the polymer self-energy $\Delta\Omega_s(r_p)$ and the local screening function $\kappa(r)$. Based on the kernel Eq.~(\ref{24}), one should note that the function $\kappa(r)$ quantifies the strength of the charge screening experienced by the polymer self-energy of Eq.~(\ref{21II}). In the inset of Fig.~\ref{fig2}(b), one notes that the counterion attraction to the DNA surface results in the interfacial charge screening excess, i.e. $\kappa(r)>\kappa_b$.  This enhances the interfacial shielding of the polymer charges and lowers the polymer free energy, giving rise to the attractive self-energy minima of the main plot. Hence, the attractive part of the interaction is due to the enhanced \textit{ionic solvation} of the polymer charges. Moreover, the larger the cation valency, the stronger the interfacial screening excess (the inset from bottom to top), and the deeper the self-energy minimum (main plot). In the presence of trivalent cations, the enhanced polymer screening overwhelms the MF-level and the image-charge induced repulsive forces, resulting in a net attractive force that drives the polymer to the DNA surface. This provides an intuitive explanation for the experimental observation of weak like-charge attraction in solutions including divalent cations and a significant attraction in the presence of trivalent cations~\cite{exp2,exp3,exp4,exp5,exp6,exp7,exp8,exp9,Cruz}. \textcolor{black}{We also note that the leading role played by the interfacial counterion abundance on the occurrence of attractive interactions and the amplification of the latter with counterion valency is in qualitative agreement with previous theoretical models with counterion-only liquids and simulations~\cite{Ha97,Ha99,jensen97}.}

\textcolor{black}{At this point, the question arises whether the like-charge polymer attraction can be observed in purely monovalent electrolytes by changing the thermodynamic characteristics of the system. In a counterion-only liquid, it is known that the importance of beyond MF charge fluctuations driving the like-charge attraction effect is characterized by the electrostatic coupling parameter $\Xi=q^2\ell_B/\mu$ where  $\mu=1/(2\pi q\ell_B\sigma_s)$ is the Gouy-Chapman length~\cite{Netz2000}. This coupling parameter scales with the characteristic system parameters as
\be\label{coup} 
\Xi\propto\frac{\sigma_d q^3}{\e_w^2T^2},
\ee
where $q$ is the counterion valency. In water solvent  with permittivity $\e_w\approx80$ at ambient temperature $T\approx 300$ K, the lowest ionic valency at which the experiments of Refs.~\cite{exp2,exp3,exp4,exp5,exp6,exp7,exp8,exp9,Cruz} observed like-charge repulsion is $q=3$. According to the scaling relation~(\ref{coup}), in order to keep the coupling strength at the same value in a monovalent solution (i.e. for $q=1$), the temperature should be reduced by a factor of five. This corresponds to the rescaled temperature $T'\approx T/5\approx60$ K located below the freezing point of water. Thus, for the polymer charges $\sigma_d$ considered in the experiments of Refs.~\cite{exp2,exp3,exp4,exp5,exp6,exp7,exp8,exp9,Cruz}, monovalent counterions cannot induce like-charge attraction except if one replaces the water solvent by another solvent of lower polarity and dielectric permittivity.}

\subsection{Analytical consideration of like-charge polymer-DNA attraction in the WC regime}
\label{sec1II}

Here, we aim to bring analytical insight into the like-charge attraction picture analyzed in Sec.~\ref{sec1}. We note that because the equations of state~(\ref{22}) and~(\ref{32}) cannot be solved analytically, the direct calculation of the polymer grand potential~(\ref{19}) is impossible.  Thus, we evaluated the polymer grand potential by introducing some approximations. This calculation is presented in Appendix~\ref{apan}. First, we restrict ourselves to the WC regime of low DNA charge or large salt where the average potential $\phi_d(r)$ is small. This regime is characterized by the inequality $\kappa_b\mu\gg1$, where $\mu=1/(2\pi\ell_B\sigma_d)$ stands for the Gouy-Chapman length corresponding to the thickness of the counterion layer around the DNA molecule. Secondly, we focus on the low multivalent ion density regime. Finally, we consider the limit of large DNA radii that can be expressed in terms of the adimensional radius $\td=\kappa_bd$ as $\td\gg1$. Within these approximations, the polymer grand potential takes the form
\bea
\label{41}
\frac{\Delta\Omega_p(\tr_p)}{L}&\approx&-\frac{\sqrt{2\pi}\tau}{s\mathrm{K}_1(\td)}\frac{e^{-\tr_p}}{\tr_p^{1/2}}+\frac{\pi\ell_B\tau^2\mathrm{I}_1(\td)}{2\mathrm{K}_1(\td)}\frac{e^{-2\tr_p}}{\tr_p}\nonumber\\
&&-\Theta\frac{\sqrt{2\pi}\ell_B\tau^2}{3s\mathrm{K}_1(\td)}\frac{e^{-\tr_p}}{\tr_p^{3/2}}.
\eea
In Eq.~(\ref{41}), we introduced the adimensional polymer position $\tr_p=\kappa_br_p$ and the parameter $s=\kappa_b\mu$ that quantifies the competition between the bulk and interfacial screening of the average potential $\phi_d(r)$. 

The first term on the right hand side of Eq.~(\ref{41}) corresponds to the MF-level polymer-DNA charge coupling. For a negatively charged polymer $\tau<0$, this term is seen to  be repulsive. The second repulsive term is induced by polymer-image charge interactions. Finally, the third term takes into account the inhomogeneous screening of the polymer charges in the vicinity of the DNA surface. This contribution involves the parameter 
\be\label{42}
\Theta=\frac{\sum_{i=1}^p\rho_{bi}q_i^3}{\sum_{i=1}^p\rho_{bi}q_i^2}
\ee
accounting for the electrolyte composition. In a symmetric electrolyte solution, $\Theta=0$. Thus, this component results from the presence of polyvalent ions that break the symmetry of the electrolyte. In the electrolyte mixture $\mbox{NaCl}+\mbox{XCl}_m$, by making use of the electroneutrality condition~(\ref{1II}), the parameter $\Theta$ can be expressed in terms of the cation densities and the polyvalent ion valency $m$ as
\be\label{43}
\Theta=(m-1)\left[1+\frac{2}{m(m+1)}\frac{\rho_{b+}}{\rho_{bm+}}\right]^{-1}.
\ee 
First of all, since $m>1$, one has $\Theta>0$. Thus, the third term of Eq.~(\ref{41}) is attractive. This unambiguously shows that like-charge attraction is induced by multivalent cations. Secondly, as $\Theta$ rises with valency ($m\uparrow\Theta\uparrow$), the larger the valency, the lower the grand potential~(\ref{41}). This explains the increase of the grand potential depth with the ionic valency in Figs.~\ref{fig2}(a) and (b). Finally, we note that in the simpler electrolyte composition $\mbox{XCl}_m$ without monovalent cations (i.e. for $\rho_{b+}=0$), Eq.~(\ref{43}) yields $\Theta=m-1$. That is, the polymer grand potential drops linearly with the ion valency. This result qualitatively agrees with the like-charged polymer interaction model of Ref.~\cite{Cruz2} where the polymer free energy was shown to scale linearly with the valency of polyvalent counterions (see Eqs. (9) and (11) of Ref.~\cite{Cruz2}). We consider next the effect of monovalent salt on like-charge polymer-DNA attraction.

\begin{figure}
\includegraphics[width=.95\linewidth]{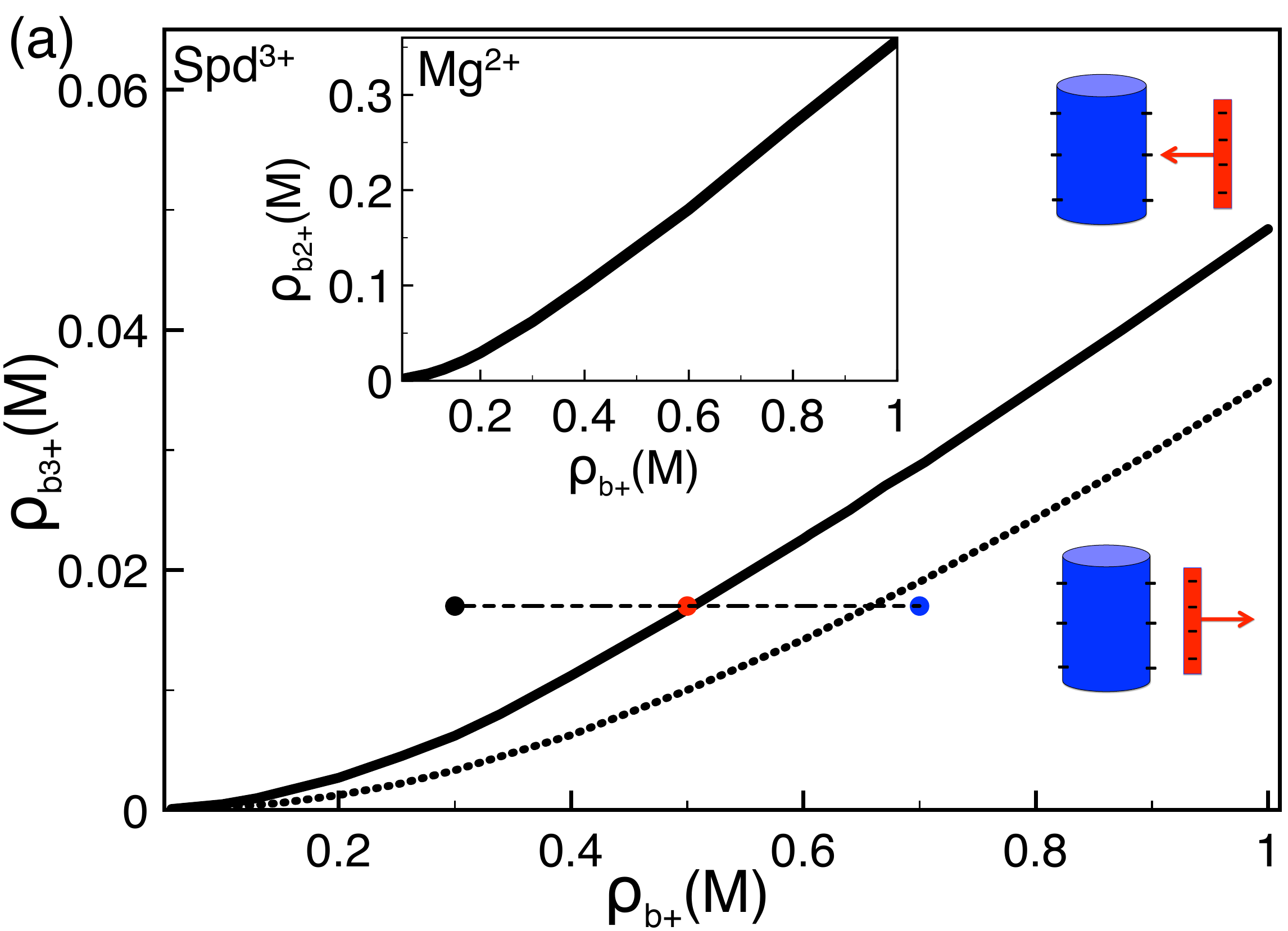}
\includegraphics[width=.95\linewidth]{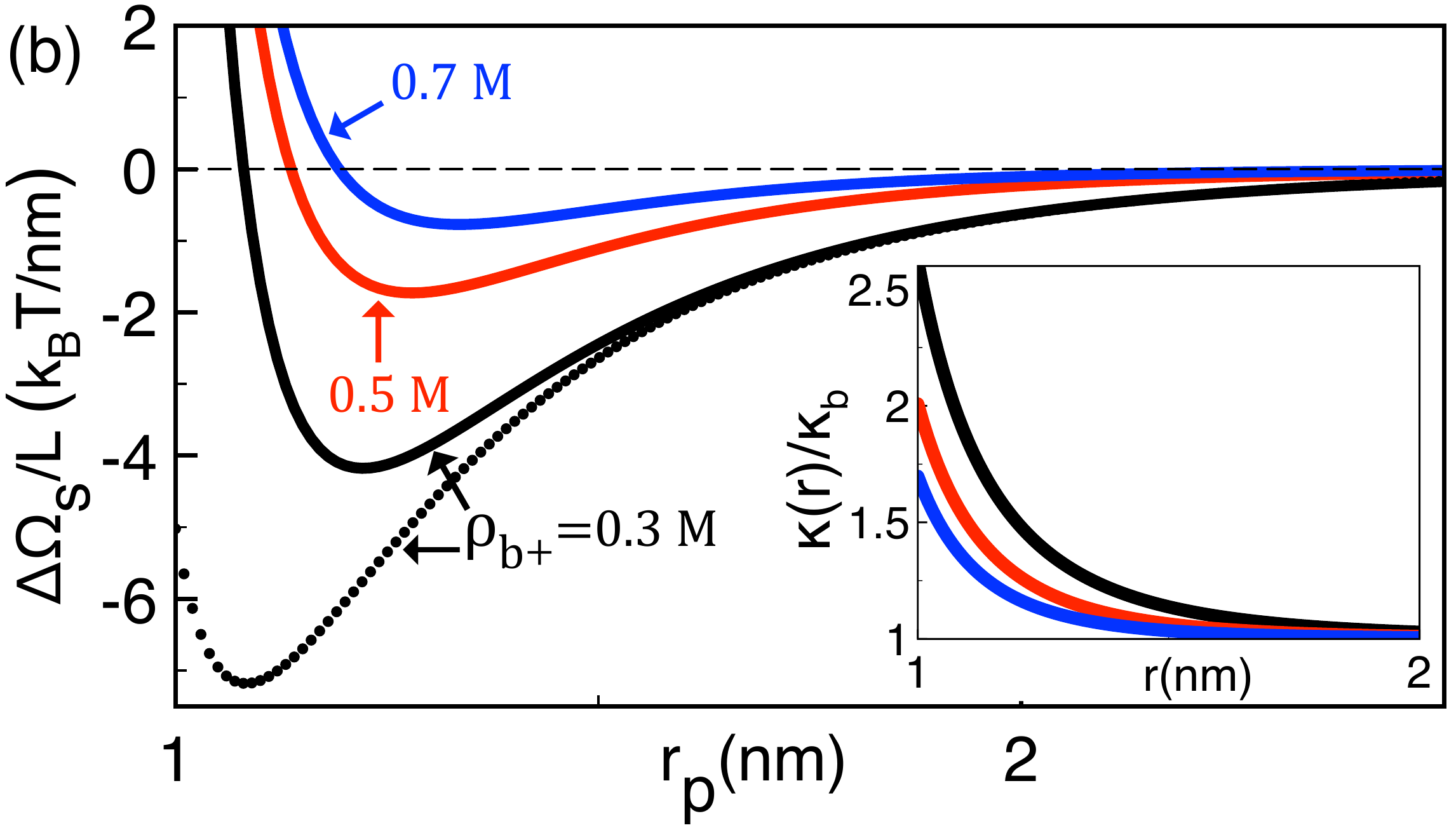}
\caption{(Color online) (a) Main plot: critical $\mbox{Spd}^{3+}$ versus $\mbox{Na}^+$ density curves for the $\mbox{NaCl}+\mbox{SpdCl}_3$ mixture separating the attractive and repulsive polymer-DNA interaction regimes (black curves). Inset: the same phase diagram for the $\mbox{NaCl}+\mbox{MgCl}_2$ liquid. (b) Self energy (main plot) and rescaled local screening function (inset) at the black dot (black curve), blue dot (blue curve), and red dot (red curve) of the phase diagram in (a). The negatively charged polymer has charge density $\tau=-0.5$ $e/${\AA}. In all figures, DNA permittivity is either $\e_d=2$ (solid curves) or $\e_d=\e_w=80$ (dotted curves).}
\label{fig3}
\end{figure}

\subsection{Suppression of like-charge attraction by monovalent counterions}
\label{sec2}

In experiments on polymer solutions, the monovalent cation density is an easily tuneable parameter. Motivated by this point, we consider here the effect of monovalent ions on the polymer-DNA interaction. The main plot of Fig.~\ref{fig3}(a) displays the critical $\mbox{Spd}^{3+}$ versus $\mbox{Na}^+$ concentrations splitting the phase domains with repulsive and attractive interactions (solid curve). The phase diagram indicates that at fixed  $\mbox{Spd}^{3+}$ density, the increment of $\mbox{Na}^+$ density turns the interaction from attractive back to repulsive. This behaviour agrees with experimental phase diagrams indicating the suppression of like-charge attraction by added monovalent cations~\cite{exp2,exp3,exp4}. Moreover, the critical curve shows that the larger is the bulk $\mbox{Na}^+$ density, the larger should be the $\mbox{Spd}^{3+}$ density in order for the like-charge attraction to survive, i.e. $\rho_{b+}\uparrow\;\rho_{b3+}\uparrow$. Such a quasilinear relation between the critical mono- and polyvalent counterion concentrations was experimentally observed in polymer solutions (see Fig.4 of Ref.~\cite{exp3} and Fig.8 of Ref.~\cite{exp4}). 

In Fig.~\ref{fig3}(b), we illustrate the mechanism behind the removal of like-charge attraction by monovalent cations. Namely, we fix the $\mbox{Spd}^{3+}$ density to $\rho_{b3+}=1.7\times10^{-2}$ M  and cross the critical line of Fig.~\ref{fig3}(a) along the dashed horizontal segment by rising the $\mbox{Na}^+$ density from $\rho_{b+}=0.3$ M to $0.7$ M. The increasing amount of $\mbox{Na}^+$ ions leads to a stronger shielding of the electrostatic potential $\phi_d(r)$. Consequently,  the cation excess at the DNA surface and the resulting interfacial screening excess are attenuated. The latter effect is shown in the inset of Fig.~\ref{fig3}(b) from top to bottom. The attenuation of the interfacial screening reduces the depth of the  self-energy well (solid curves in the main plot) and cancels the overall attraction.  Thus, monovalent cations remove like-charge attraction by weakening the enhanced solvation of the polymer charges in the vicinity of the DNA surface. This effect is also seen in the attractive component of the grand potential in Eq.~(\ref{41}). According to Eq.~(\ref{43}), the amplitude of the attractive component decreases with the rise of the monovalent cation density, i.e. $\rho_{b+}\uparrow\Theta\downarrow$. The functional form of Eq.~(\ref{43}) indicates that in order for the like-charge attraction to survive, the increment of the monovalent cation density has to be compensated by a stronger $\mbox{Spd}^{3+}$ concentration, which explains the positive slope of the critical line in Fig.~\ref{fig3}(a). 

To evaluate the effect of the cation valency on the phase diagram of Fig.~\ref{fig3}(a), in the inset, we reported the cation density range where like-charge attraction is expected in the  $\mbox{NaCl}+\mbox{MgCl}_2$ liquid. The comparison with the main plot shows that at given  $\mbox{Na}^+$ concentration, the minimum  $\mbox{Mg}^{2+}$ density for the occurrence of like-charge attraction is almost an order of magnitude higher than the critical $\mbox{Spd}^{3+}$ density of the $\mbox{NaCl}+\mbox{SpdCl}_3$ mixture. This point confirms again the absence of significant like-charge polymer attraction in electrolyte mixtures including divalent cations at physiological concentrations. 

\begin{figure}
\includegraphics[width=.95\linewidth]{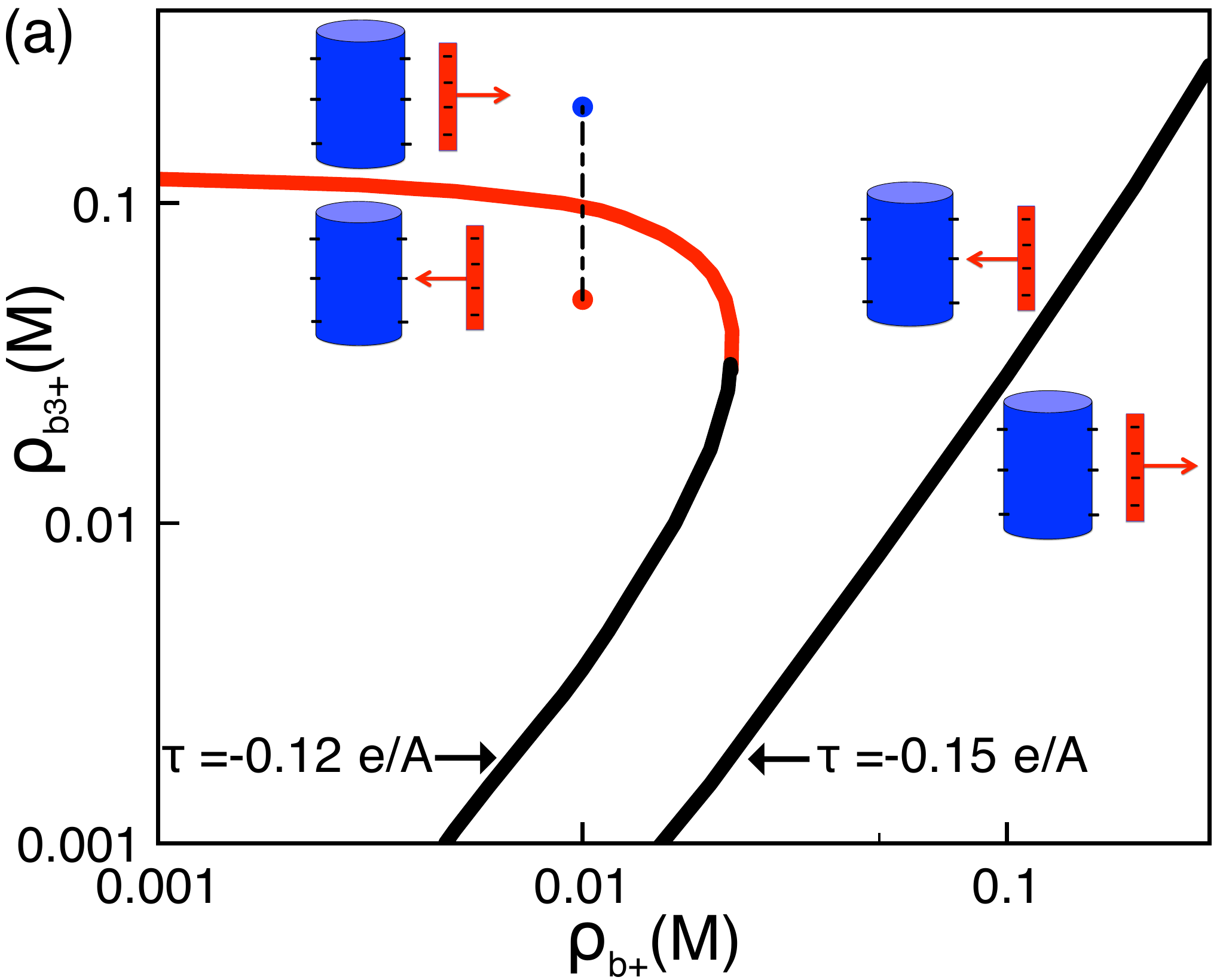}
\includegraphics[width=.95\linewidth]{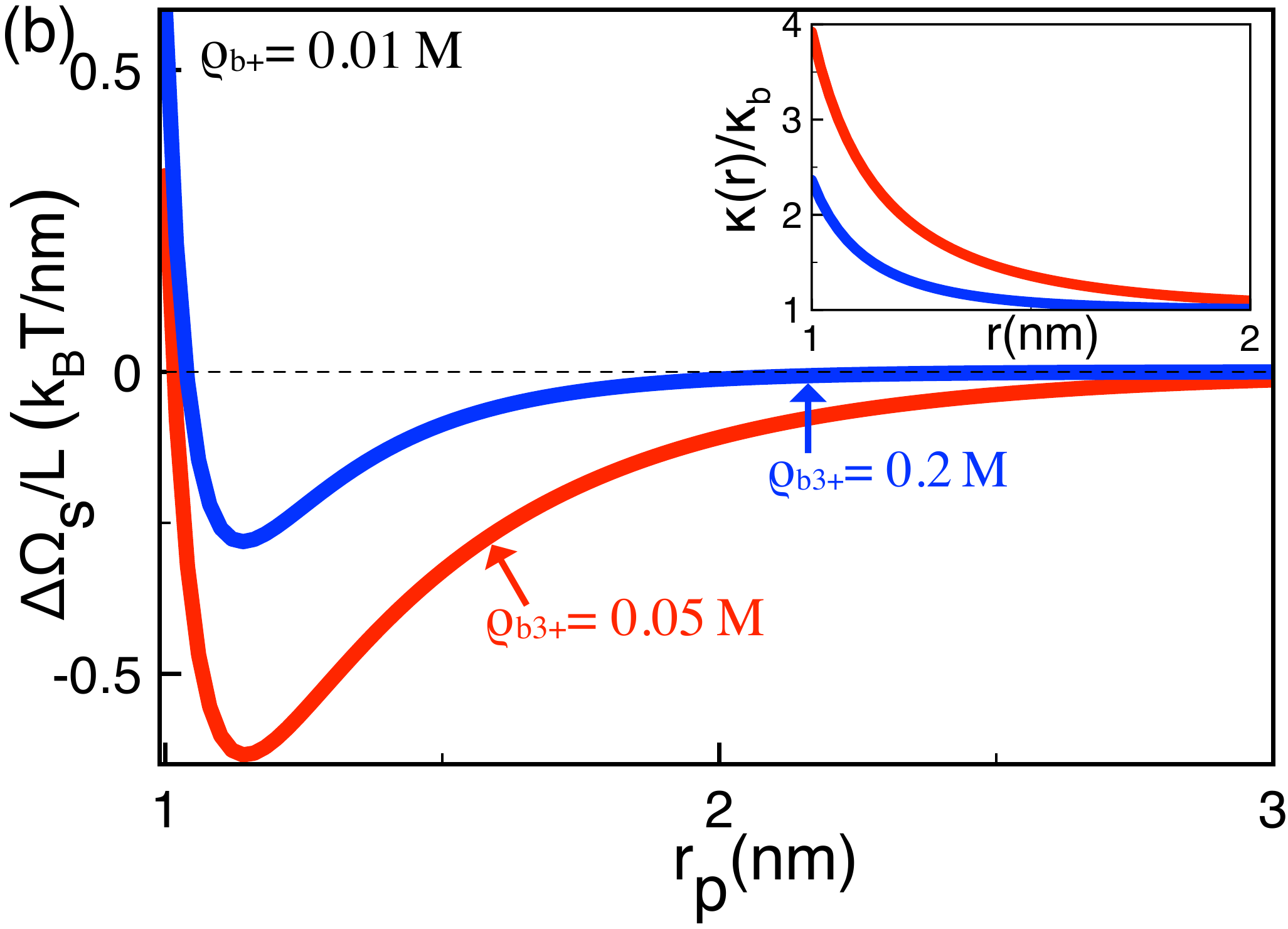}
\caption{(Color online) (a) Critical $\mbox{Spd}^{3+}$ versus $\mbox{Na}^+$ concentration curves splitting the regimes with attractive and repulsive polymer-DNA interaction at two different polymer charge densities. Figure (b) illustrates the polymer self-energy~(\ref{21II}) (main plot) and the local screening function~(\ref{25}) (inset) at the red dot (red curve) and blue dot (blue curve) of the phase diagram in (a). The remaining parameters are given in the caption of Fig.~\ref{fig1}.}
\label{fig3II}
\end{figure}

The variation of the salt density can equally favour water binding to the DNA molecule and increase its dielectric permittivity~\cite{DNAwater}.  In order to scrutinize the impact of DNA hydration on like-charge attraction, we reproduced the critical density and self-energy plots of Figs.~\ref{fig3}(a) and (b) by turning off the dielectric contrast (dotted curves for $\e_d=\e_w$). Comparing the self-energy curves at $\rho_{b+}=0.3$ M, one notes that this limit removes the repulsive image charge barrier and deepens the self-energy minimum.  Due to the enhanced attraction, at the larger DNA permittivity $\e_d=\e_w$, the polymer binding occurs at lower $\mbox{Spd}^{3+}$ concentrations (compare the dotted and solid curves in Fig.~\ref{fig3}(a)). Thus, DNA hydration favours like-charge attraction. That being said, the moderate difference between the critical curves with and without image forces indicates that the dielectric contrast does not play a major role on the like-charge polymer attraction. This peculiarity is due to the different range of the electrostatic forces experienced by the polymer. Indeed, Eq.~(\ref{41}) shows that the image-charge component of the grand potential scaling as $\sim e^{-2\tr_p}$ is shorter ranged than the MF and the attractive components decaying as $\sim e^{-\tr_p}$ with the polymer distance. This point qualitatively agrees with recent MC simulations of like-charged polymer-membrane interactions where the effect of the membrane polarization on the polymer binding was shown to be minimal~\cite{Levin3}. 
 
 \subsection{Effect of polymer charge and reentrant dissolution}
 \label{sec3}

In this part, we investigate the effect of the polymer charge on polymer-DNA interactions. In Fig.~\ref{fig3II}(a), we reproduced the phase diagram of Fig.~\ref{fig3}(a) in a weaker polymer charge density regime.  At the charge density $\tau=-0.15$ $e/${\AA}, the critical $\mbox{Spd}^{3+}$  concentration rises steadily with the bulk $\mbox{Na}^+$ density, i.e. $\rho_{b+}\uparrow\rho_{b3+}\uparrow$. As seen in Fig.~\ref{fig3}(a), this monotonous behaviour scrutinized in Sec.~\ref{sec2} survives for stronger polymer charges.  However, for polymers with the charge density below the value $|\tau|\approx0.15$ $e/${\AA}, this behaviour is not monotonous. At the polymer charge density $\tau=-0.12$ $e/${\AA}, the critical line of Fig.~\ref{fig3II}(a) includes a second branch (red curve) where the critical $\mbox{Spd}^{3+}$ concentration decreases with the $\mbox{Na}^+$ density, i.e.  $\rho_{b+}\uparrow\rho_{b3+}\downarrow$. The black and red branches of the critical line join at the characteristic $\mbox{Na}^+$ concentration $\rho_{b+}=0.02$ M beyond which the like-charge attraction regime ceases to exist, regardless of how strong is the $\mbox{Spd}^{3+}$  concentration. Next, we elucidate this peculiarity by considering the physics behind the red branch of the critical curve. 

In Fig.~\ref{fig3II}(a), the non-uniform trend of the critical line for $\tau=-0.12$ $e/${\AA} has an important implication. Starting at the bottom of the figure and crossing vertically the black and red branches by rising the $\mbox{Spd}^{3+}$  density, the polymer-DNA interaction turns respectively from repulsive to attractive and back to repulsive. The corresponding reentrant behaviour has been observed in previous experiments where multivalent cations were found to induce polymer aggregation (i.e. attraction) at low densities and redissolution (repulsion) at large densities~\cite{exp2,exp3,exp5,exp7,exp8,exp9,Cruz}. The counterion-induced polymer binding mechanism behind the first regime was investigated in Sec.~\ref{sec1}. We consider now the redissolution regime characterized by the red branch of the critical curve. In Fig.~\ref{fig3II}(b), we plotted the local screening function and the polymer self-energy at the red and blue dots of the phase diagram. In this high density regime, the MF-level shielding of the DNA-induced potential $\phi_d(r)$ by $\mbox{Spd}^{3+}$  molecules comes into play. As a result, rising  the $\mbox{Spd}^{3+}$ density along the dashed vertical line of the phase diagram, the cation binding to DNA and the associated screening excess are attenuated (see the inset of Fig.~\ref{fig3II}(b)). This diminishes the depth of the attractive self-energy well (main plot) and switches the polymer-DNA interaction from attractive back to repulsive.  Hence, at large concentrations, polyvalent cations behave similar to monovalent salt and remove like-charge polymer attraction by suppressing charge correlations. 

\begin{figure}
\includegraphics[width=1.0\linewidth]{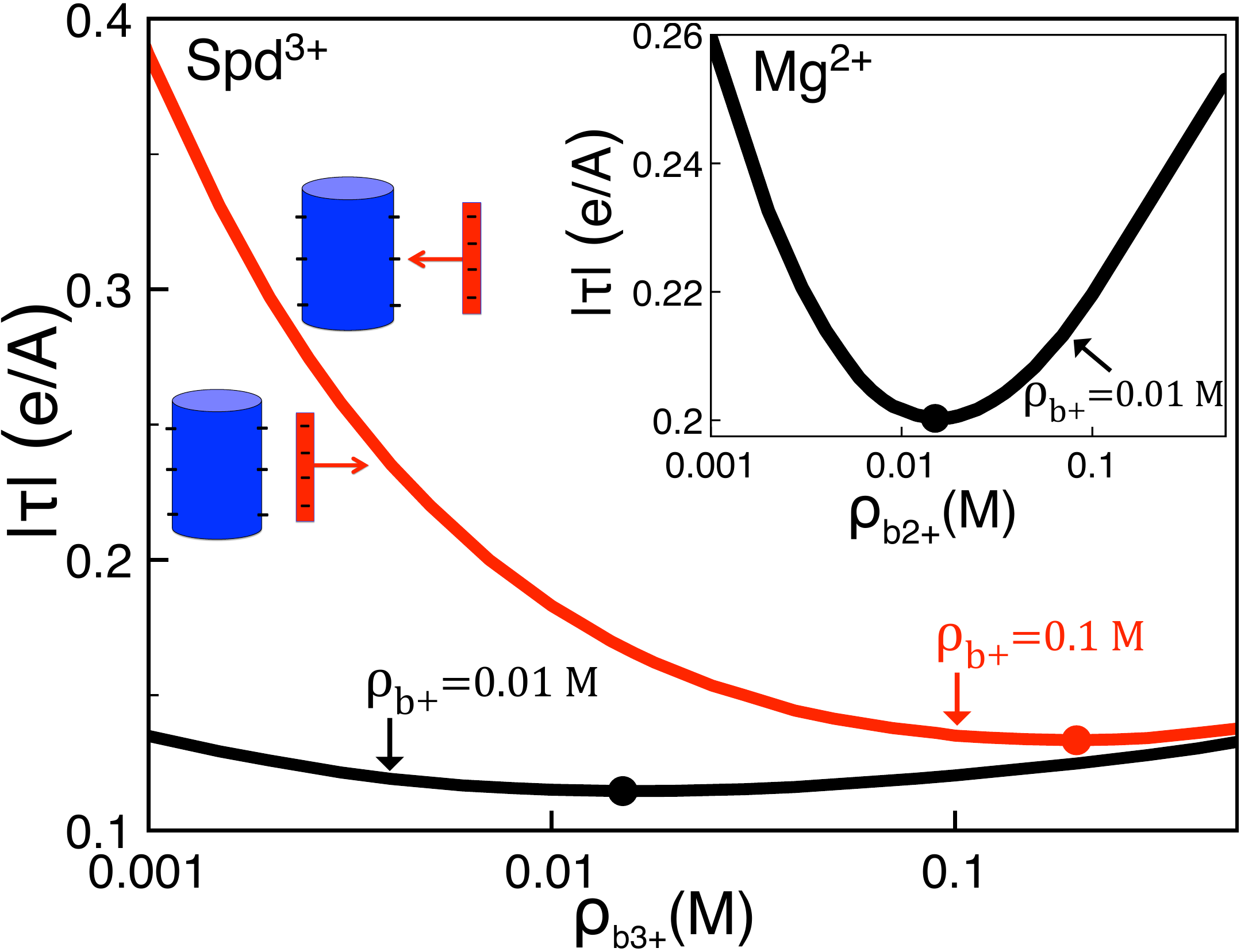}
\caption{(Color online) Characteristic polymer charge against $\mbox{Spd}^{3+}$  density of the $\mbox{NaCl}+\mbox{SpdCl}_3$ mixture splitting the attractive and repulsive interaction regimes. The inset displays the same phase diagram for the $\mbox{NaCl}+\mbox{MgCl}_2$ liquid. The remaining parameters are the same as in Fig.~\ref{fig1}.}
\label{fig4}
\end{figure}

We investigate now the influence of the polymer charge strength on the like-charge polymer attraction. Fig.~\ref{fig4} displays the evolution of the critical polymer charge with the $\mbox{Spd}^{3+}$ density at two different $\mbox{Na}^+$ concentrations. The non-monotonous behaviour of the critical curves is a consequence of the reentrant behaviour scrutinized above. Namely, the progressive addition of $\mbox{Spd}^{3+}$ molecules to the $\mbox{NaCl}$ solution initially lowers the critical polymer charge, i.e. $\rho_{b3+}\uparrow\;\tau\downarrow$.  This regime is characterised by the strengthening of the attractive polymer self-energy by $\mbox{Spd}^{3+}$ molecules.  In order for the like-charge attraction to persist, its attenuation by a weaker polymer charge has to be compensated by a larger amount of $\mbox{Spd}^{3+}$ molecules. This explains the negative slope of the critical curve. Rising the $\mbox{Spd}^{3+}$ density beyond the minimum of the curve,  one gets into the reentrant regime where $\mbox{Spd}^{3+}$ molecules screen the average potential $\phi_d(r)$ and weaken the interfacial screening excess.  In order for the interaction to remain attractive, this effect has to be overwhelmed by a stronger polymer charge. This leads to the rise of the critical curve ($\rho_{b3+}\uparrow\;\tau\uparrow$).

A key information provided by the phase diagram of Fig.~\ref{fig4} is the location of the turning point; this corresponds to the weakest polymer charge density $\tau^*$ where like-charge attraction can be observed. Comparing the critical lines of the main plot, one notes that the critical polymer charge drops with the $\mbox{Na}^+$ concentration ($\rho_{b+}\downarrow\;\tau\downarrow$) but the value of $\tau^*$ is weakly sensitive to the amount of monovalent cations. Indeed, in the physiological $\mbox{Na}^+$ concentration regime $0.01\;\mbox{M}\leq\rho_{b+}\leq0.1\;\mbox{M}$, the minimum polymer charge density stays in the range $\tau^*\approx0.12-0.14\;e/\mbox{{\AA}}$. However, the critical polymer charge is significantly sensitive to the cation valency; the inset shows that in the $\mbox{NaCl}+\mbox{MgCl}_2$ liquid, both the critical charge $\tau$ and its lower bound $\tau^*$ are twice as large as their corresponding values in the $\mbox{Spd}^{3+}$ liquid.

\begin{figure}
\includegraphics[width=1.1\linewidth]{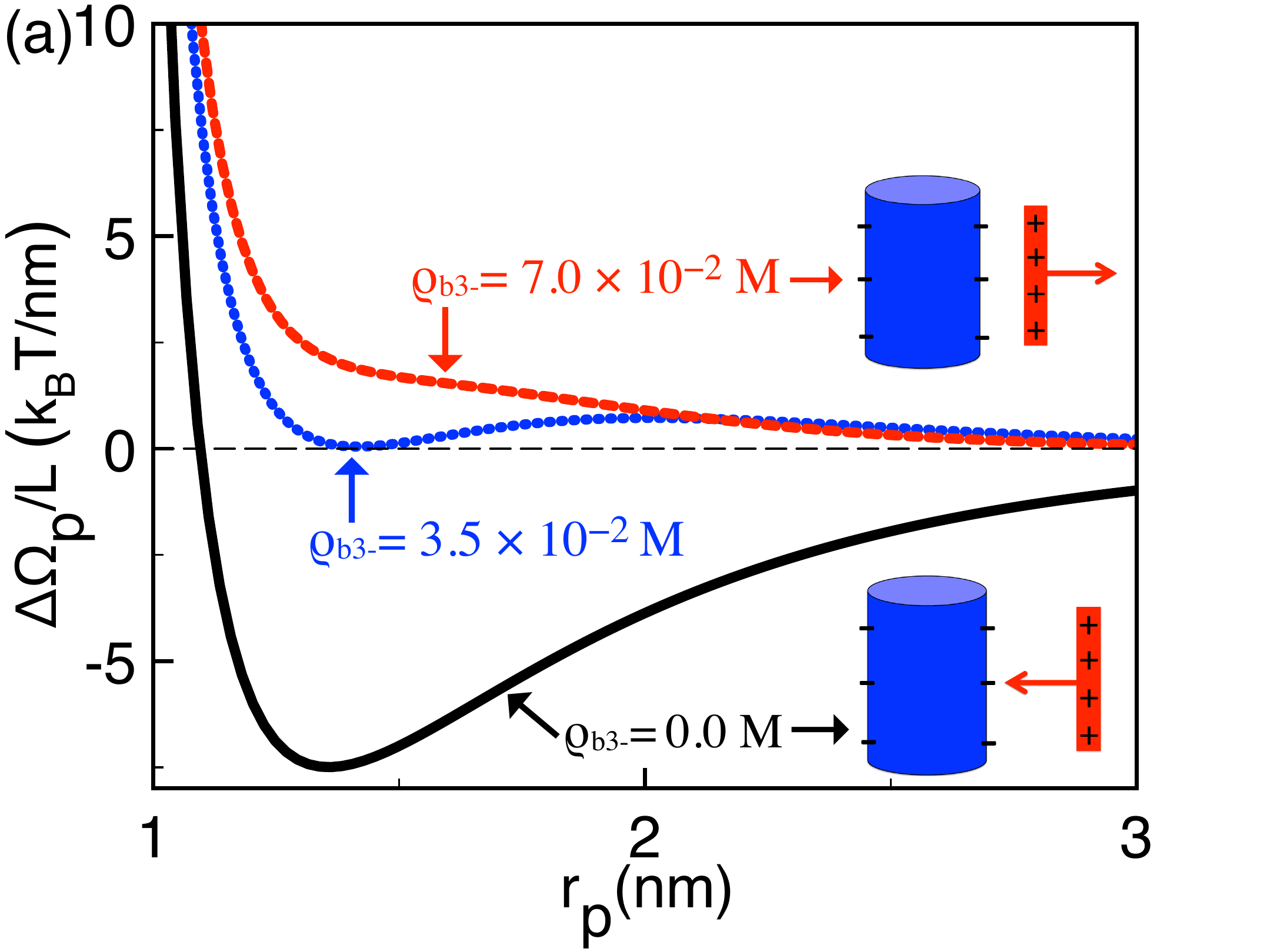}
\includegraphics[width=1.1\linewidth]{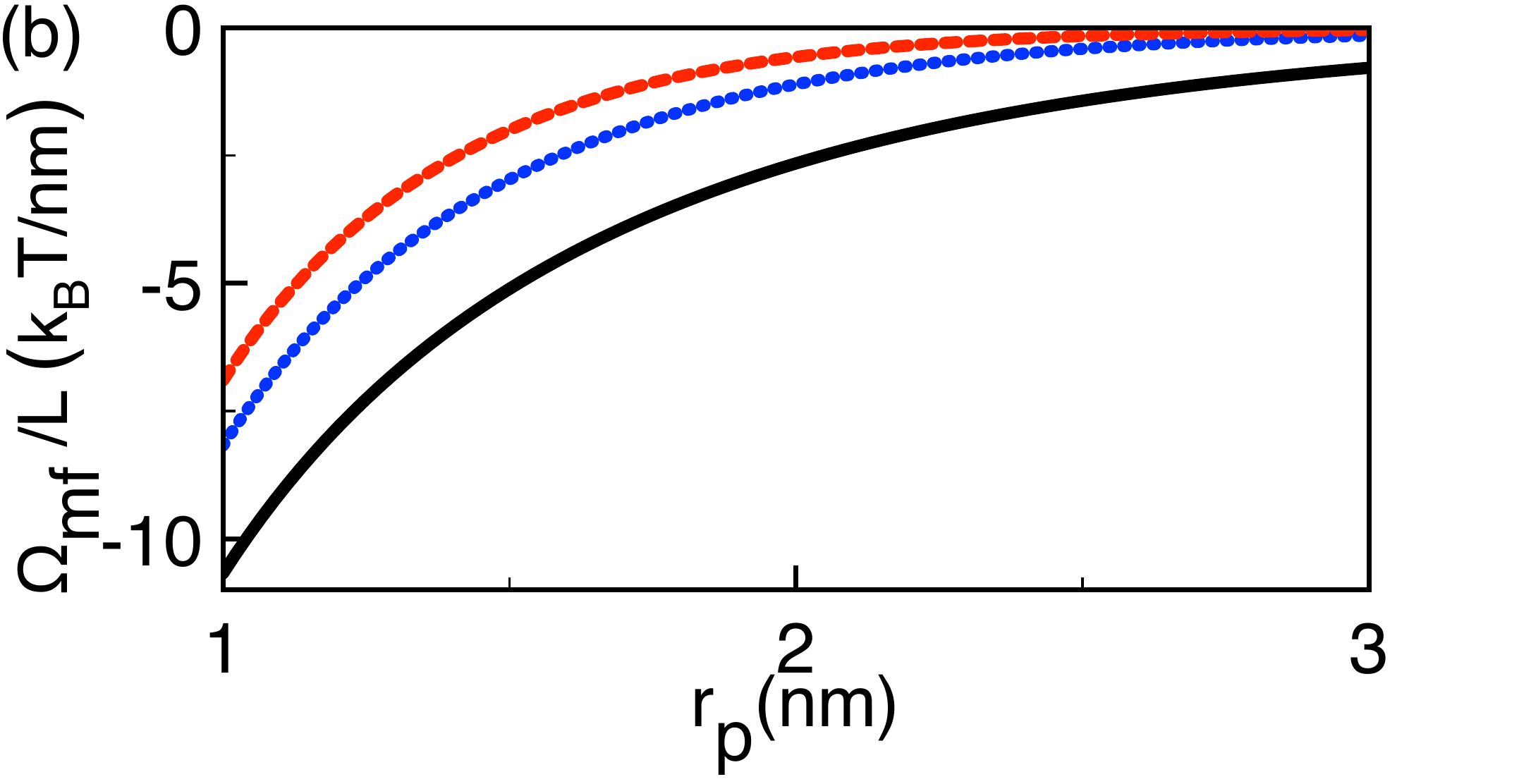}
\includegraphics[width=1.1\linewidth]{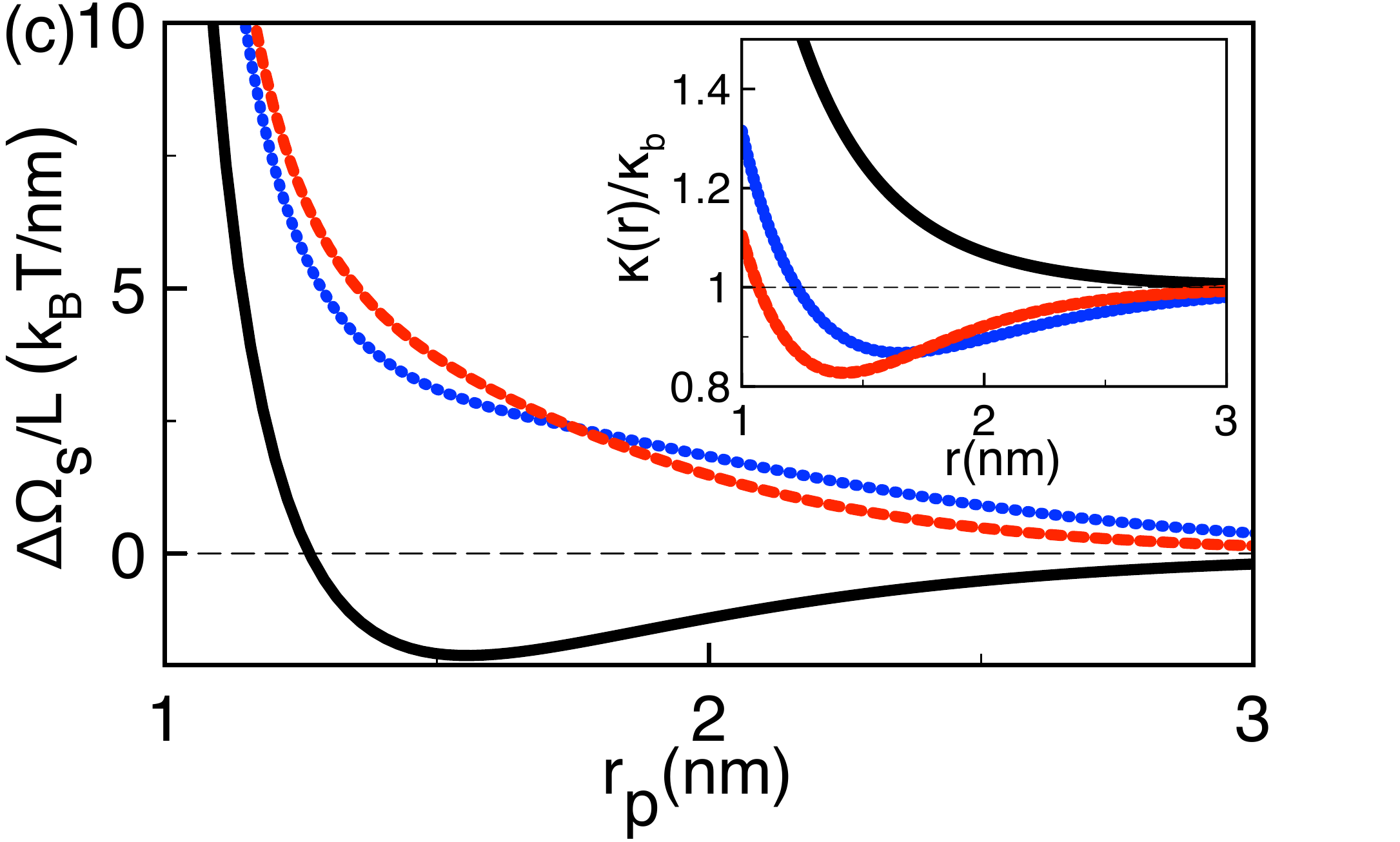}
\caption{(Color online) (a) Total polymer grand potential~(\ref{19}), (b) MF grand potential~(\ref{20}), (c) self-energy~(\ref{21II}) (main plot) and the local screening function~(\ref{25}) (inset). The positively charged polymer has charge density $\tau=0.5$ $e/${\AA}. In the electrolyte mixture $\mbox{NaCl}+\mbox{PO}_4\mbox{Na}_3$, the bulk $\mbox{Cl}^-$ concentration is set to $\rho_{b-}=0.1$ M. $\mbox{PO}^{3-}_4$ density is $\rho_{b3-}=0.0$ M (black curves), $3.5\times10^{-2}$ M (blue curves), and $7.0\times10^{-2}$ M (red curves). The remaining parameters are given in the caption of Fig.~\ref{fig1}.}
\label{fig5}
\end{figure}

\section{Cationic polymer-DNA repulsion}
\label{rep}

This section is devoted to the interaction between the DNA molecule and a positively charged polyelectrolyte ($\tau>0$). The polymer-DNA complex is immersed in the electrolyte mixture  $\mbox{NaCl}+\mbox{PO}_4\mbox{Na}_3$ composed of monovalent $\mbox{Na}^+$  and $\mbox{Cl}^-$ ions, and trivalent phosphate anions $\mbox{PO}^{3-}_4$. In the following part where we will vary the anion densities $\rho_{b-}$ and $\rho_{b3-}$, the $\mbox{Na}^+$ concentration will be set by the electroneutrality condition of Eq.~(\ref{1II}) that reads $\rho_{b+}=\rho_{b-}+3\rho_{b3-}$. In Sec.~\ref{op1}, an opposite-charge repulsion mechanism induced by polyvalent anions is investigated. The effect of the DNA charge strength on the opposite-charge decomplexation  is scrutinized in Sec.~\ref{op2}.

\subsection{Opposite-charge polymer-DNA repulsion mechanism}
\label{op1}

Fig.~\ref{fig5}(a) illustrates the grand potential density $\Delta\Omega_p/L$ of the polymer immersed in the electrolyte mixture $\mbox{NaCl}+\mbox{PO}_4\mbox{Na}_3$. The curves correspond to various $\mbox{PO}^{3-}_4$ concentrations. The $\mbox{Cl}^-$ density is set to $\rho_{b-}=0.1$ M. In the $\mbox{NaCl}$ liquid (black curve), the interaction is characterized by an attractive well. The well is induced by the MF-level opposite-charge attraction embodied in the MF potential $\Omega_{mf}$ (see Fig.~\ref{fig5}(b)), and the interfacial image-charge barrier brought by the self-energy $\Delta\Omega_s$ (see Fig.~\ref{fig5}(c)). Adding $\mbox{PO}^{3-}_4$ ions of bulk density $\rho_{b3-}=3.5\times10^{-2}$ M, the grand potential rises and the attractive minimum becomes bistable (blue curve in Fig.~\ref{fig5}(a)). At the larger $\mbox{PO}^{3-}_4$ concentration $\rho_{b3-}=7.0\times10^{-2}$ M (red curve), the minimum disappears and the polymer grand potential becomes purely repulsive. Thus, in the presence of a sufficient amount of polyvalent anions, the positively charged polymer is repelled by the negatively charged DNA molecule. This opposite-charge decomplexation effect is one of the key results of our article.

In order to understand the mechanism driving the opposite-charge polymer repulsion, we focus on the MF and self-energy components of the polymer grand potential. Fig.~\ref{fig5}(b) shows that the increase of the $\mbox{PO}^{3-}_4$ density simply reduces the amplitude of the attractive MF grand potential. This effect is due to the stronger shielding of the DNA potential $\phi_d(r)$ by a larger amount of salt. We consider now the polymer self-energy (the main plot of Fig.~\ref{fig5}(c)) and the rescaled screening function (the inset) that should be interpreted together. In the $\mbox{NaCl}$ solution (black curves), the $\mbox{Na}^+$ condensation at the DNA surface results in the interfacial screening excess $\kappa(r)>\kappa_b$, which leads to the weakly attractive self-energy well.  As the $\mbox{PO}^{3-}_4$ ions are added to the solution, they are repelled by the DNA charges and driven to the bulk region. Due to their high valency, their depletion from the DNA surface results in the interfacial charge screening deficiency with respect to the bulk, i.e. $\kappa(r)<\kappa_b$ (blue and red curves). This means that the bulk electrolyte can more efficiently screen the polymer charges, which translates into a purely repulsive self-energy (main plot). Furthermore, one sees that the larger the bulk $\mbox{PO}^{3-}_4$ density, the stronger the interfacial screening deficiency, and the more repulsive the self-energy at the DNA surface.  As the bulk $\mbox{PO}^{3-}_4$ density exceeds the characteristic value $\rho_{b3-}=3.5\times10^{-2}$ M, the repulsion induced by the interfacial $\mbox{PO}^{3-}_4$ exclusion becomes strong enough to overcome the MF-level attraction.  This results in the unbinding of the cationic polymer from the DNA molecule. We finally note that  this polyvalent anion-induced decomplexation can be also explained in terms of the WC grand potential of Eq.~(\ref{41}). Indeed, for a general electrolyte mixture $\mbox{NaCl}+\mbox{XNa}_m$ including the polyvalent anions $X^{m-}$, the amplitude of the correlation-correction term given by Eq.~(\ref{42}) is
\be\label{44}
\Theta=-(m-1)\left[1+\frac{2}{m(m+1)}\frac{\rho_{b-}}{\rho_{bm-}}\right]^{-1}.
\ee
Since $\Theta<0$, correlations associated with polyvalent anions bring a repulsive contribution to the polymer grand potential. Next, we consider the effect of the DNA charge strength on this polyvalent anion-induced opposite-charge repulsion mechanism.

\subsection{Effect of DNA charge strength}
\label{op2}

\begin{figure}
\includegraphics[width=1.05\linewidth]{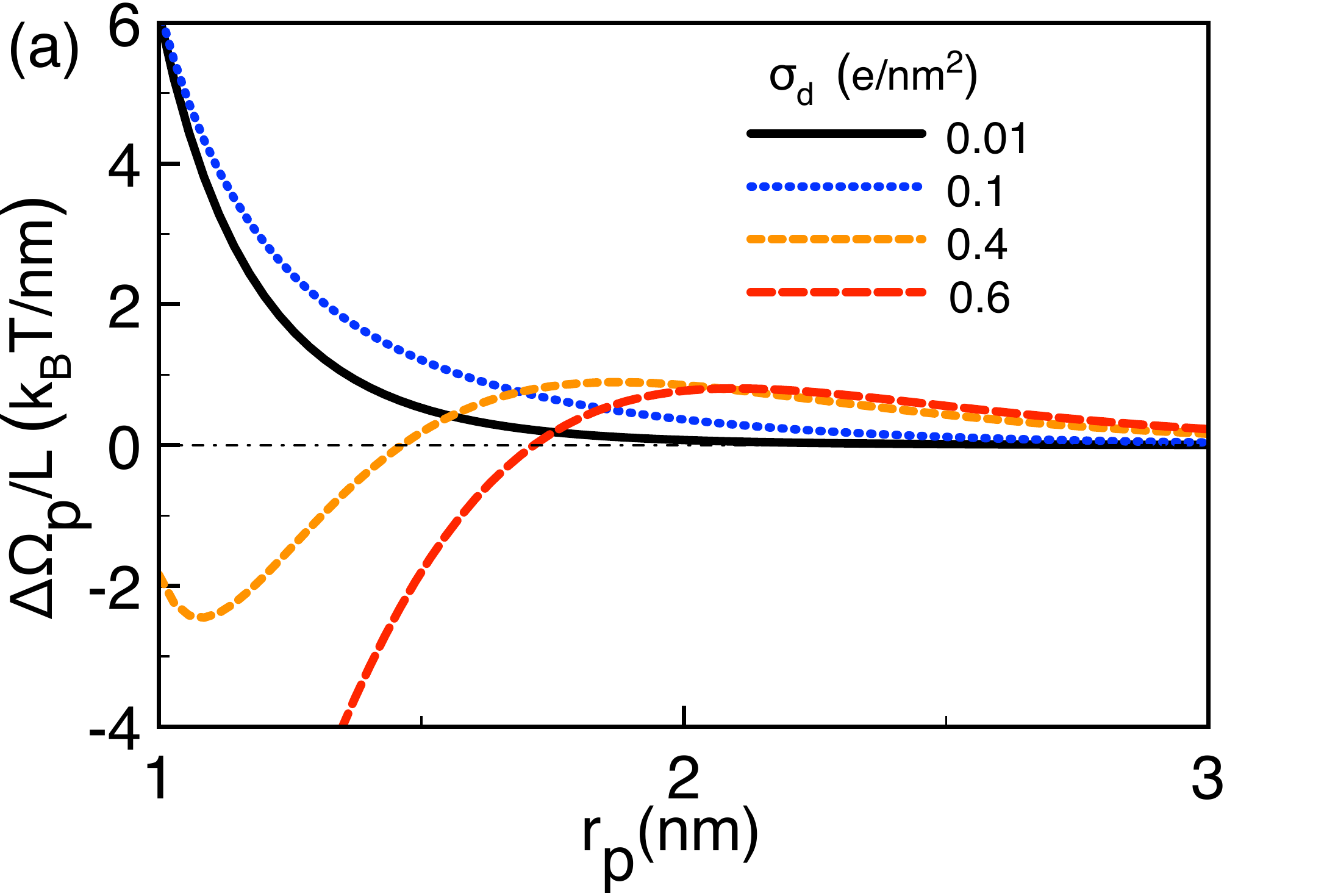}
\includegraphics[width=1.05\linewidth]{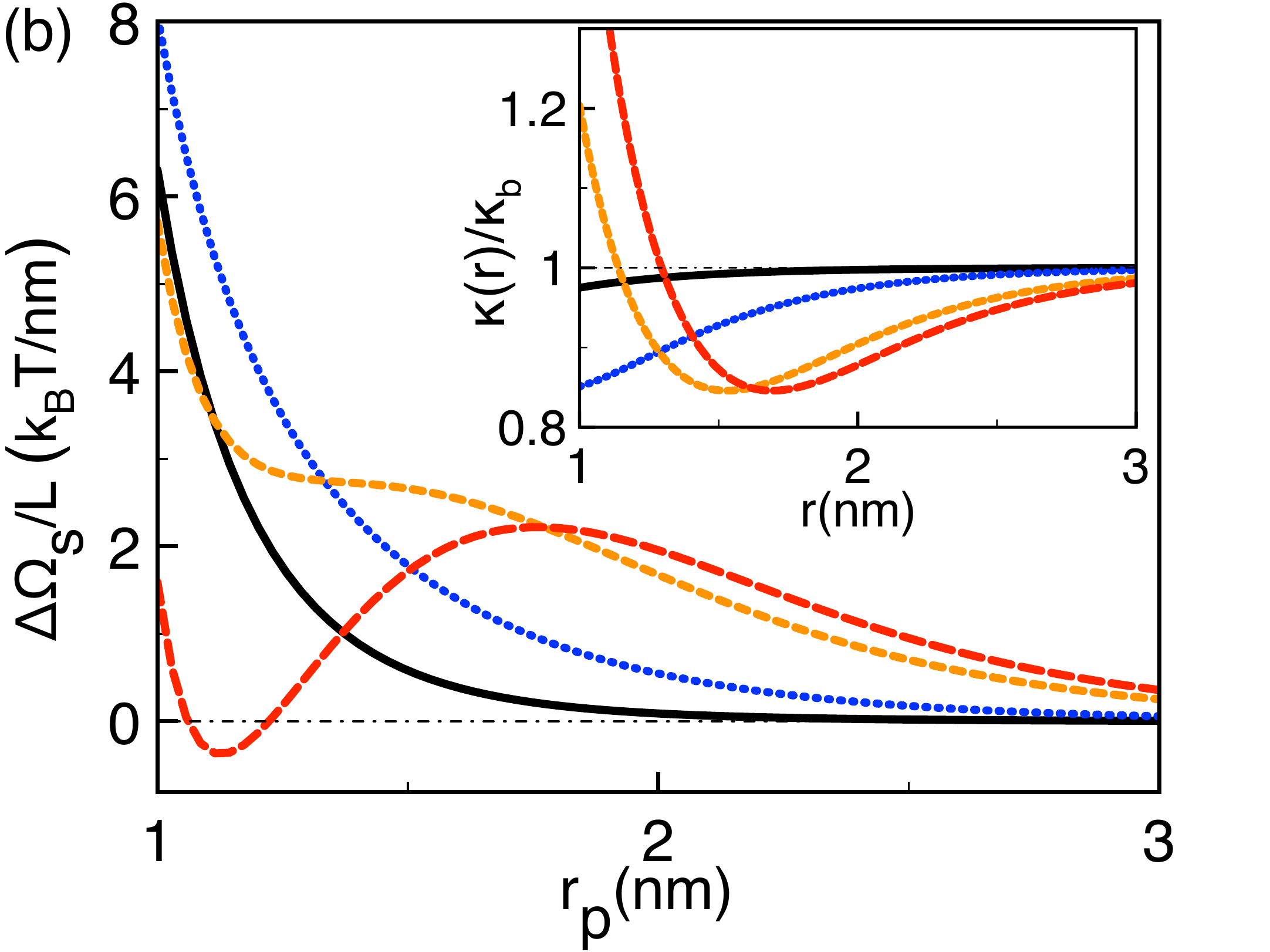}
\caption{(Color online) (a) Total polymer grand potential~(\ref{19}) and (b) self-energy~(\ref{21II}) (main plot). The positively charged polymer has charge density $\tau=0.5$ $e/${\AA}. In the electrolyte mixture $\mbox{NaCl}+\mbox{PO}_4\mbox{Na}_3$, the bulk $\mbox{Cl}^-$ and $\mbox{PO}^{3-}_4$  concentrations are respectively $\rho_{b-}=0.1$ M and $\rho_{b3-}=5\times10^{-2}$ M. The inset of (b) displays the local screening function~(\ref{25}). The remaining parameters are given in the caption of Fig.~\ref{fig1}.}
\label{fig6}
\end{figure}

In the presence of polyvalent anions, we found that the DNA surface charge gives rise to two competing effects : the direct charge-charge coupling resulting in the opposite-charge attraction and the multivalent anion depletion that induces the opposite-charge repulsion. Thus, the question arises as to how the overall polymer-DNA interaction is influenced by the rise of the DNA charge density. To investigate this point, in Fig.~\ref{fig6}(a), we plotted the polymer grand potential at various DNA charge densities. Rising the DNA charge from $\sigma_d=0.01$ $e/\mbox{nm}^2$ (black curve) to  $\sigma_d=0.1$ $e/\mbox{nm}^2$ (blue curve), the total grand potential becomes more repulsive. This results from the self-energy component of the grand potential. The former is displayed in Fig.~\ref{fig6}(b) (main plot) together with the rescaled screening function (inset). The comparison of the black and blue curves shows that in this regime, the rise of the DNA charge amplifies the  $\mbox{PO}^{3-}_4$ exclusion and the resulting interfacial screening deficiency, i.e. $\sigma_d\uparrow\kappa(r)\downarrow$. As a result, the self-energy barrier rises with the DNA charge density.

Fig.~\ref{fig6}(a) shows that at the larger charge densities  $\sigma_d=0.4$ $e/\mbox{nm}^2$ (orange curve) and  $0.6$ $e/\mbox{nm}^2$ (red curve), the polymer grand potential becomes more repulsive for $r_p\gtrsim2.0$ nm but  strongly drops and acquires an attractive well  below this distance. Indeed, in this higher charge regime, a larger DNA surface charge not only amplifies the $\mbox{PO}^{3-}_4$  exclusion but also gives rise to the interfacial $\mbox{Na}^+$ excess. In the inset of Fig.~\ref{fig6}(b),  one sees that this peculiarity results in the charge screening excess $\kappa(r)>\kappa_b$ in the vicinity of the DNA surface and charge screening deficiency $\kappa(r)<\kappa_b$ outside the interfacial region. Consequently, upon the increase of the DNA charge, the polymer self-energy weakly rises far from the DNA surface but strongly drops and becomes attractive in the interfacial region. This non-uniform behaviour explains the grand potential turnover from repulsive to attractive in Fig.~\ref{fig6}(a). Thus, a strong enough macromolecular charge suppresses the  opposite-charge repulsion effect.

\section{Conclusions}

In this work, we investigated electrostatic polymer-DNA interactions in electrolyte mixtures. In Sec.~\ref{subsec1}, we developed a general test charge approach that allows to calculate the electrostatic grand potential of a weakly charged body interacting with a macromolecule of arbitrary charge strength. The approach developed therein extends to electrolyte mixtures the formalism introduced in Ref.~\cite{PRE2016} for symmetric electrolytes. Within this test charge theory, in Sec.~\ref{subsec2}, we calculated the grand potential of a polymer interacting with a DNA molecule. The grand potential is composed of two components. The MF interaction term takes into account the direct DNA-polymer charge coupling. This term embodies the MF-level like-charge repulsion and opposite-charge attraction effects. The second grand potential component is the polymer self-energy that breaks the MF interaction picture. Namely, this component brings correlation-corrections associated with the ionic cloud deformation induced by the DNA charges as well as image-charge forces resulting from the low permittivity of the DNA molecule. The evaluation of the self-energy necessitates the solution of the non-uniformly screened electrostatic Green's equation in cylindrical coordinates. In order to achieve this task, we introduced a numerical inversion scheme that allows to solve this equation by iteration.

Sec.~\ref{at} was devoted to the interaction of a negatively charged polymer with a DNA molecule. The addition of polyvalent cations into a monovalent electrolyte results in the like-charged polymer-DNA attraction.  The mediator of the like-charge attraction is the dense counterion layer developed around the DNA molecule, which acts as an enhanced screening environment and lowers the polymer free energy.  This effect is reversed at large multivalent cation densities. In this regime, the MF-level shielding of the DNA-induced potential by multivalent counterions comes into play. As a result, multivalent cations weaken the interfacial cation excess and turn the interaction from attractive back to repulsive.  Due to the same MF-level screening mechanism, the increment of monovalent cations systematically suppresses the like-charge polymer attraction. The above-mentioned features are in qualitative agreement with experiments on like-charged polymer solutions~\cite{exp2,exp3,exp4,exp5,exp6,exp7,exp8,exp9,Cruz}.

Within the same formalism, in Sec.~\ref{rep}, we considered the opposite situation of a positively charged polymer interacting with a DNA molecule. Therein, we identified a new polyvalent anion-induced opposite-charge repulsion mechanism. The effect is driven by the repulsion of the polyvalent anions by the DNA charges. The resulting anion depletion from the DNA surface weakens the screening ability of the interfacial region with respect to the bulk electrolyte. This induces a repulsive force that results in the decomplexation of the positively charged polymer from the DNA molecule.  This is the key prediction of our article.  We also scrutinized the role played by the DNA charge strength on this opposite-charge decomplexation effect. In addition to the polyvalent anion exclusion, the DNA charges lead to monovalent cation excess at the surface. Beyond a characteristic surface charge, the cation excess overcompensates the anion depletion. Thus, in this region, the rise of the macromolecular charge weakens the interfacial charge screening deficiency and cancels the opposite charge decomplexation. 

We introduced \textcolor{black}{the first} unified theory of like-charge attraction and opposite-charge repulsion between polyelectrolytes and DNA molecules in electrolyte mixtures. Our formalism is based on electrostatics and neglects some features of these highly complex systems. First of all, we neglected the conformational fluctuations of the interacting polyelectrolytes. This simplification was motivated by the experimental evidence that polymer conformations play no qualitative role in multivalent cation-induced polymer aggregation~\cite{exp4}. That being said, it should be noted that a field theoretic model unifying polymer fluctuations and electrostatic interactions was ingeniously developed by Tsonchev et al. in Ref.~\cite{dun}. This approach may be a promising way to integrate polymer fluctuations into our formalism in a future work, though the tremendous complexity of this unified theory is beyond the scope of our article. \textcolor{black}{At this point, we note that the first extension of our theory in this direction would be the inclusion of the polymer orientation. In a future work, we plan to add this complication to our model in order to investigate correlation effects on the diffusion-driven regime of polymer capture by membrane nanopores.} The second approximation of our theory is the test charge approach that neglects the reaction of the electrolyte to the linear polymer. Our motivation behind the test charge approximation was explained in the Introduction : the full 1l level consideration of the polymer charges breaks the cylindrical symmetry of the problem and the model becomes analytically intractable. Finally,  our theory is based on the 1l formulation of charge interactions. In a future work, the underlying 1l description can be extended by using electrostatic formulations able to cover the parameter regime from weak to strong electrostatic coupling~\cite{SCHatlo}. However, it should be noted that this improvement will considerably increase the numerical complexity of the solution scheme and shadow the transparency of the physical picture emerging from our simpler formalism. 

In order to determine the validity regime of our theory at a quantitative level, comparisons with numerical simulations are needed. To our knowledge, simulation results for the present linear polymer-cylindrical DNA model are not available in the literature. As a first step to test the quantitative accuracy of our theory, we currently  work on the simulation of a simpler system : an anionic polymer interacting with a like-charged membrane that we recently modelled in Ref.~\cite{PRE2016}. At the next step, we plan to extend this numerical scheme to the more complicated case of ions and polymers surrounding the cylindrical DNA molecule. \textcolor{black}{Indeed, as discussed in the Introduction part, different theoretical approaches have been so far used to scrutinize correlation effects on polymer condensation. In order to evaluate comparatively the quantitative accuracy of these models and also the consequences of our approximations on the physical conclusions made in our work, a systematic comparison with numerical simulations is clearly needed. Since continuum theories of charged systems cannot yet account for various complications included in Molecular Dynamics simulations such as solvent charge structure and finite ion size, we believe that such a comparison should be made by reducing the complexity level of the simulated models to the level of the theories.} This being said, we emphasize that despite the above-mentioned approximations, our theory can qualitatively reproduce several characteristics of like-charge polymer attraction observed in experiments. This point indicates that our model is able to capture \textcolor{black}{qualitatively} the essential physics driving the like-charge attraction phenomenon. Moreover, our new prediction of the opposite-charge decomplexation effect can be verified by experiments or simulations.  Finally, our theoretical predictions may provide guiding information for gene delivery techniques and genetic engineering methods where polymer-DNA interactions play a key role.

\smallskip
\appendix
\section{Calculating the polymer grand potential in the weak-coupling regime}
\label{apan}

In this appendix, we present the analytical evaluation of the polymer grand potential~(\ref{19}) in the weak coupling (WC) regime of low macromolecular charges. We start with the calculation of the MF grand potential~(\ref{20}). For weak surface charges or strong salt, the average potential $\phi_d(r)$ is small. Thus, linearizing the PB Eq.~(\ref{22}), one obtains
\be\label{ap1}
\frac{1}{r}\partial_r\left[r\partial_r\phi_d(r)\right]-\kappa_b^2\theta(r-d)\phi_d(r)=4\pi\ell_B\sigma_d\delta(r-d).
\ee
The solution to this equation reads
\be\label{ap2}
\phi_d(r)=-\frac{2}{s}\frac{\mathrm{K}_0(\kappa_br)}{\mathrm{K}_1(\kappa_bd)},
\ee
where we introduced the parameter $s=\kappa_b\mu$ with the Gouy-Chapman length $\mu=1/(2\pi\ell_B\sigma_d)$. Substituting the potential~(\ref{ap2}) into Eq.~(\ref{20}), the MF-level polymer grand potential takes the form
\be
\label{ap3}
\Omega_{mf}(r_p)=-\frac{2L\tau}{s}\frac{\mathrm{K}_0(\kappa_br_p)}{\mathrm{K}_1(\kappa_bd)}.
\ee

The evaluation of the self-energy component~(\ref{21II}) involving the Green's function $\tv_m(r,r';0)$ is non-trivial.  First, by using the integral relation~(\ref{32}) together with the DH Green's function in Eq.~(\ref{39}), the self-energy~(\ref{21II}) can be expressed as
\be
\label{ap4}
\Delta\Omega_s(r_p)=L\ell_B\tau^2\sum_{m-\infty}^{+\infty}\left[F_m(0)\;\mathrm{K}^2_m(\kappa_br_p)+u_m(r_p)\right],
\ee
where we defined the potential
\be
\label{ap5}
u_m(r_p)=\frac{1}{4\pi\ell_B}\int_d^\infty\mathrm{d}rr\;\tv_{0,m}(r_p,r;0)\delta n(r)\tv_m(r,r_p;0).
\ee
The integral in Eq.~(\ref{ap5}) cannot be evaluated analytically. To progress further,  we Taylor-expand the ion density excess function $\delta n(r)$ defined by Eq.~(\ref{33}) in terms of the average potential $\phi_d(r)$. Furthermore, we note that within the same WC approach, correlation corrections brought by the density excess function $\delta n(r)$ are small. Thus, we restrict ourselves to the first iterative solution of Eq.~(\ref{32}). This is equivalent to replacing in Eq.~(\ref{ap5}) the Green's function $\tv_m(r,r_p;0)$ by its WC limit $\tv_{0,m}(r,r_p;0)$. Consequently, Eq.~(\ref{ap5}) takes the form
\be
\label{ap6}
u_m(r_p)=\sum_{i=1}^p\frac{\rho_{bi}q_i^3}{4\pi\ell_B}\int_d^\infty\mathrm{d}rr\;\tv_{0,m}(r_p,r;0)\phi_d(r)\tv_{0,m}(r,r_p;0).
\ee
Now, we substitute into Eq.~(\ref{ap6}) the average potential in Eq.~(\ref{ap2}) and the DH Green's function in Eq.~(\ref{39}). Then, we pass to the adimensional coordinates by defining the rescaled radial variables $\tr=\kappa_b r$ and $\tr_p=\kappa_b r_p$, and the rescaled pore length $\td=\kappa_bd$. The potential~(\ref{ap6}) takes the form
\bea
\label{ap7}
u_m(\tr_p)=-\frac{2\Theta}{s\mathrm{K}_1(\td)}\left[I_m(\tr_p)+J_m(\tr_p)\right]
\eea
In Eq.~(\ref{ap7}), we introduced the auxiliary parameter $\Theta=\sum_{i=1}^p\rho_{bi}q_i^3/\sum_{i=1}^p\rho_{bi}q_i^2$ and the adimensional functions
\bea
\label{ap8}
I_m(\tr_p)&=&\int_{\td}^{\tr_p}\mathrm{d}\tr\tr\;\mathrm{K}_0(\tr)\\
&&\hspace{5mm}\times\left\{\mathrm{I}_m(\tr)\mathrm{K}_m(\tr_p)+F_m(0)\mathrm{K}_m(\tr)\mathrm{K}_m(\tr_p)\right\}^2\nonumber\\
\label{ap9}
J_m(\tr_p)&=&\int_{\tr_p}^{\infty}\mathrm{d}\tr\tr\;\mathrm{K}_0(\tr)\\
&&\hspace{5mm}\times\left\{\mathrm{I}_m(\tr_p)\mathrm{K}_m(\tr)+F_m(0)\mathrm{K}_m(\tr_p)\mathrm{K}_m(\tr)\right\}^2.\nonumber
\eea
From now on, we will restrict ourselves to large macromolecular radii or strong salt $\td\gg1$ where the ground state mode $m=0$ brings the major contribution to the self-energy. Thus, in Eq.~(\ref{ap4}), we will neglect the components with finite index $|m|>0$. Furthermore, considering the inequality $\tr_p>\td\gg1$, we will evaluate the integrals in Eqs.~(\ref{ap8}) and~(\ref{ap9}) by replacing the modified Bessel functions of the integrands with their large distance limit~\cite{math},
\be\label{ap10}
\mathrm{I}_0(x)\approx\frac{e^x}{\sqrt{2\pi x}}\;;\hspace{1cm}\mathrm{K}_0(x)\approx\sqrt{\frac{\pi}{2x}}e^{-x}\hspace{5mm}\mathrm{for}\;x\gg1.
\ee
Noting that $F_0(0)=\mathrm{I}_1(\td)/\mathrm{K}_1(\td)$, within the above-mentioned approximations, the self-energy Eq.~(\ref{ap4}) takes the closed-form
\be\label{ap11} 
\Delta\Omega_s(\tr_p)\approx L\ell_B\tau^2\frac{\mathrm{I}_1(\td)}{\mathrm{K}_1(\td)}\mathrm{K}^2_0(\tr_p)-\frac{2L\ell_B\tau^2}{s\mathrm{K}_1(\td)}\Theta\Psi(\tr_p),
\ee
with the adimensional function
\bea\label{ap12}
\Psi(\tr_p)&=&\frac{\mathrm{K}_0^2(\tr_p)}{6\sqrt2}\left\{3\left[\mathrm{Erfi}(\sqrt{\tr_p})-\mathrm{Erfi}(\sqrt{\td})\right]\right.\\
&&\hspace{1.2cm}+6\pi F_0(0)\left[\mathrm{Erf}(\sqrt{\tr_p})-\mathrm{Erf}(\sqrt{\td})\right]\nonumber\\
&&\hspace{1.2cm}\left.+\sqrt{3}\pi^2 F_0^2(0)\left[\mathrm{Erf}(\sqrt{3\tr_p})-\mathrm{Erf}(\sqrt{3\td})\right]\right\}\nonumber\\
&&-\frac{\pi^2}{2\sqrt{6}\;\mathrm{K}_1^2(\td)}\left[-1+\mathrm{Erf}(\sqrt{3\tr_p})\right]\nonumber\\
&&\hspace{1.7cm}\times\left[\mathrm{I}_1(\td)\mathrm{K}_0(\tr_p)+\mathrm{K}_1(\td)\mathrm{I}_0(\tr_p)\right]^2.\nonumber
\eea
In Eq.~(\ref{ap12}), we made used of the error function $\mathrm{Erf}(x)$ and its imaginary counterpart $\mathrm{Erfi}(x)=\mathrm{Erf}(ix)/i$~\cite{math}. Finally, in order to be consistent with the preceding calculation, we evaluate the asymptotic large distance behaviour of the grand potential components~(\ref{ap3}) and~(\ref{ap11}). Namely, we expand these components by using the asymptotic limit of the Bessel functions in Eq.~(\ref{ap10}), and the error function
\be\label{ap13}
\mathrm{Erf}(x)\approx1-\frac{e^{-x^2}}{\sqrt{\pi}x}\hspace{5mm}\mathrm{for}\;x\gg1.
\ee
Substituting the simplified results into Eq.~(\ref{19}), after some algebra, one gets the total polymer grand potential Eq.~(\ref{41}) of the main text.

\end{document}